\documentclass[epj]{svjour}

\usepackage{graphicx,amssymb,amsmath}

\newcommand{\be}{\begin{equation}}
\newcommand{\ee}{\end{equation}}

\newcommand{\bea}{\begin{eqnarray}}
\newcommand{\eea}{\end{eqnarray}}

\newcommand{\br}{\mathbf{r}}

\newcommand{\bn}{\mathbf{n}}

\newcommand{\bq}{\mathbf{q}}
\newcommand{\bB}{\mathbf{B}}
\newcommand{\bM}{\mathbf{M}}

\newcommand{\tnabla}{\tilde{\nabla}}
\def\eq#1{eq.~(\ref{#1})}
\def\eqs#1#2{eqs.~(\ref{#1},\ref{#2})}

\usepackage[toc,page]{appendix}
\usepackage{array}
\newcolumntype{C}[1]{>{\centering\let\newline\\\arraybackslash\hspace{0pt}}m{#1}}
\newcolumntype{L}[1]{>{\raggedright\let\newline\\\arraybackslash\hspace{0pt}}m{#1}}
\newcolumntype{R}[1]{>{\raggedleft\let\newline\\\arraybackslash\hspace{0pt}}m{#1}}

\begin{document}

\def\myequation{\stepcounter{equation}\(\displaystyle }
\def\endmyequation{\hfill \hbox{\enspace(\theequation)}\)}

\title{Mixed lipid bilayers with locally varying spontaneous curvature and bending}

\author{Guillaume Gueguen \and Nicolas Destainville \and Manoel Manghi\thanks{Corresp. author, \email{manghi@irsamc.ups-tlse.fr}}}

\institute{Universit\'e de Toulouse, UPS, Laboratoire de Physique Th\'eorique (IRSAMC), F-31062 Toulouse, France, EU \and
CNRS; LPT (IRSAMC); F-31062 Toulouse, France, EU}

\date{\today}

\abstract{
A model of lipid bilayers made of a mixture of two lipids with different average compositions on both leaflets, is developed. A Landau hamiltonian describing the lipid-lipid interactions on each leaflet, with two lipidic fields $\psi_1$ and $\psi_2$, is coupled to a Helfrich one, accounting for the membrane elasticity, via both a local spontaneous curvature, which varies as $C_0+C_1(\psi_1-\psi_2)/2$, and a bending modulus equal to $\kappa_0+\kappa_1(\psi_1+\psi_2)/2$. This model allows us to define curved patches as membrane domains where the asymmetry in composition, $\psi_1-\psi_2$, is large, and thick and stiff patches where $\psi_1+\psi_2$ is large. These thick patches are good candidates for being lipidic rafts, as observed in cell membranes, which are composed primarily of saturated lipids forming a liquid-ordered domain and are known to be thick and flat nano-domains.
The lipid-lipid structure factors and correlation functions are computed for globally spherical membranes and planar ones and for a whole set of parameters including the surface tension and the coupling in the two leaflet compositions. Phase diagrams are established, within a Gaussian approximation, showing the occurrence of two types of Structure Disordered phases, with correlations between either curved or thick patches, and an Ordered phase, corresponding to the divergence of the structure factor at a finite wave vector. The varying bending modulus plays a central role for curved membranes, where the driving force $\kappa_1C_0^2$ is balanced by the line tension, to form raft domains of size ranging from 10 to 100~nm. For planar membranes, raft domains emerge \textit{via} the cross-correlation with curved domains. A global picture emerges from curvature-induced mechanisms, described in the literature for planar membranes, to coupled curvature- and bending-induced mechanisms in curved membranes forming a closed vesicle.}

\PACS{
      {PACS-87.16.D-}{Membranes, bilayers, and vesicles}   \and
      {PACS-87.16.dt}{Structure, static correlations, domains, and rafts}   \and
      {PACS-82.70.Uv}{Surfactants, micellar solutions, vesicles, lamellae, amphiphilic systems, (hydrophilic and hydrophobic interactions)}
     }

\maketitle

\section{Introduction}

Biological plasma membranes are fluid mosaics made of several thousands different types of lipids and proteins, necessary for the cell to modulate its local membrane composition to achieve the various biological functions (e.g. cellular signal transduction, and trafficking with either the cytosol or the inter-cellular medium). Among the various heterogeneous structures of the plasma membrane, the concept of lipid rafts has emerged more than 20 years ago~\cite{simons1,simons2,pike,jacobson}. These nanoscopic domains are assumed to be platforms of sorting and signalisation by recruiting specific proteins. These rafts have a lipidic composition enriched in sphingolipids and cholesterol which induces a liquid ordered phase thicker than the surrounding membrane. 

Although this concept of raft is still under debate, it has been connected to the observation of lipid-lipid phase separation in model multicomponent bilayers~\cite{mouritsen,veatch1,veatch2,feigenson,baumgart}. For instance, it has been observed very recently in giant unilamellar vesicles the formation of nano-domains and modulated phases in mixtures of various phosphatidylcholines (DSPC, DOPC and POPC) and cholesterol~\cite{konyakhina1,goh}. The experimental parameter that tune the various modulated phases is the fraction of DOPC lipids. Moreover coarse-grained molecular dynamics simulations of binary lipid mixtures in a flat monolayer~\cite{stevens,perlmutter} have shown the formation of nano thicker gel domains. These domains have also been observed in numerical simulations of binary bilayers~\cite{bagatolli,schmid}.

Since the seminal works by Leibler and Andelman~\cite{leibler1,leibler2}, it has been shown that introducing a linear coupling between the local curvature and the lipidic composition --which means that one lipid species tends to curve the membrane-- leads to the formation of finite size domains for planar mixed lipidic monolayers and bilayers~\cite{mackintosh,schick}. Indeed, for large coupling the homogeneous (or disordered) phase becomes unstable at a critical wave vector $q^*$, at which the structure factor diverges, which leads to the formation of modulated (or ordered) stripe or hexagonal phases~\cite{sunilkumar}. This phase transition is usually preceded by a homogeneous but structured phase where correlations between lipids exist which reflects a tendency towards order (maximum of the structure factor at $q^*$). This regime of liquid structured on a length $\xi \equiv \sqrt{\kappa_0/\sigma}$, where $\kappa_0$ is the membrane bending modulus and $\sigma$ its surface tension, is consistent with the nano-domain, or ``raft'', sizes of 10 to 100~nm, for cell membrane elastic parameter values~\cite{schick,shlomovitz}. A slightly different model, where the coupling is introduced between the lipid composition and a field related to the lipidic unsaturated tail orientation, leads to similar conclusions~\cite{brewster,hirose}. Indeed by replacing this field by $\nabla h$ (where $h$ is the height of the membrane), the model is equivalent to the local curvature-mediated one. 
\begin{figure}[!t]
\centering
\includegraphics[width=.35\textwidth]{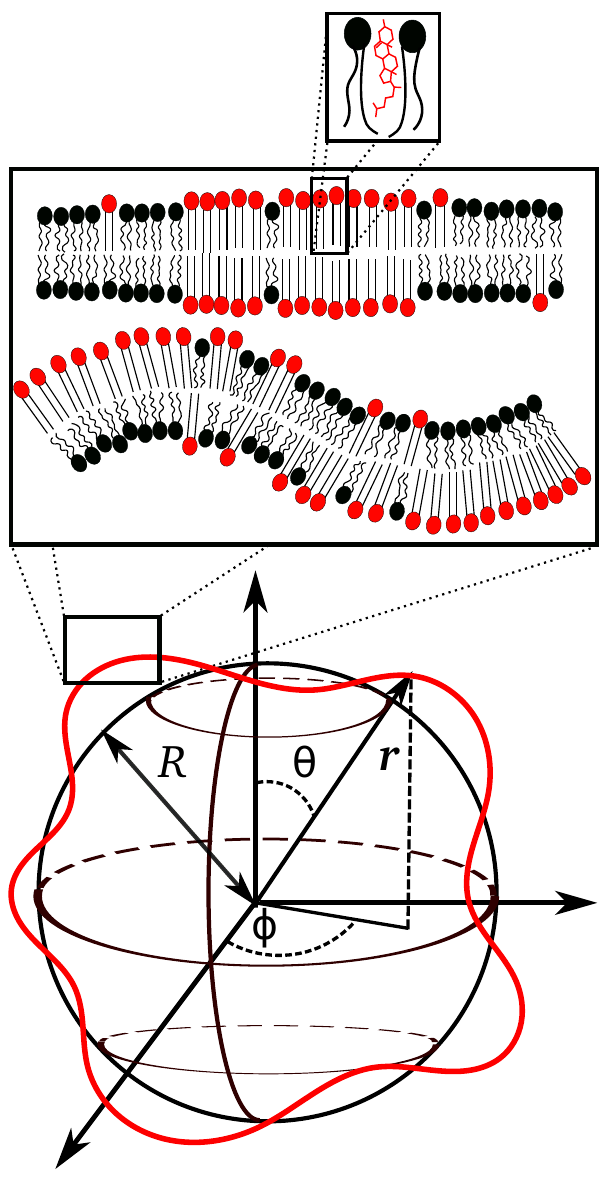}
\caption{\label{fig1} Sketch of a fluctuating vesicle around a reference sphere of radius $R$ and definition of the spherical coordinates. The insets show thick and curved patches induced by different local composition in lipids of type A (red) and B (black) in the bilayer membrane. A thicker patch and thus a locally larger bending modulus is due to an excess of lipid A in both leaflets (top) and a local curvature is due to different lipid compositions in the two leaflets (bottom). The curved lipids A can be sen as a model of sphingolipids with cholesterols (in red) inserted between them.}
\end{figure}
However these curvature-mediated models for bilayers considered only planar membranes with a vanishing averaged spontaneous curvature, which we shall prove to be a drastic restriction. They did not consider the most general and much richer case of curved membranes forming a closed vesicle. 

In this paper, we develop a more general model, where the membrane is considered to be possibly curved, infinitely thin and composed of a bilayer made of a mixture of two types of lipids A and B. According to the relative local composition in lipids in the two monolayers, either curved patches or thick patches may appear, as sketched in fig.~\ref{fig1}. Since the thick patches are stiffer, they are described in the theory through a composition dependent bending modulus, following the model of ref.~\cite{dean} for planar lipidic monolayers, a similar model having also been developed for describing the elasticity of DNA close to its denaturation~\cite{DNA}. In~\cite{dean}, the authors show that the dependence of the bending modulus on 
the composition does not play any role at the Gaussian level, but by considering a cumulant expansion in the height field, a micro-phase can appear. We will also introduce below a new term in the Hamiltonian that depends on the average asymmetry of lipid compositions between both leaflets, and we show it plays a pivotal role.
These thick patches, which are symmetric and rich in A-type of lipids, are thus good candidates for lipid rafts with the condition that a composite ``lipid'' A should be viewed as a sphingolipid associated to a cholesterol as sketched in fig.~\ref{fig1} (in a strict sense, we should study a three-component bilayer, but the mathematics would then be virtually intractable).
Contrary to models where rigid inclusions in the membrane are treated as boundary conditions for the membrane height~\cite{goulian,fournier,evans}, the thicker patches are treated with a third field which is the composition locally averaged on the two leaflets. The local spontaneous curvature is associated to the difference between the composition of the two leaflets, as in real cells~\cite{lodish}, and both fields are coupled to the height fluctuations. In refs.~\cite{safran1990,safran1991,mac1993,taniguchi1994}, the spontaneous curvature is also dependent on the difference between the two monolayer compositions, but our Hamiltonian is written differently. In this model first proposed by Safran and collaborators, each monolayer is frustrated by its own spontaneous curvature, which depends on composition, whereas in our case, both leaflets of the vesicle are stressed identically. We treat the general case of an almost spherical membrane that lead to \textit{bending-mediated mechanism even at the Gaussian level}, which disappears in the limit of planar membranes.

The paper is structured as follows. The general model is presented in Section~II. The correlation functions of the curving and stiffening fields, which are a common way to characterise modulated phases (see, e.g., \cite{fan2010}), are computed in Section~III. The case of a planar membrane is recovered in Section~IV in the limit of infinite vesicle radius and the phase diagram is computed analytically. In Section~V, are described the structured disordered and ordered phases for a spherical membrane. In particular, it is shown that contrary to the planar case, the bending-induced mechanism favors the formation of ordered phase of rafts even for a low coupling between the two monolayers. In Section~VI, our theoretical results are compared to previous ones for bi-component planar membranes, and we discuss to what extent our model, in which experimental values for the elastic parameters are injected, can explain the nano-domains observed in biological membranes or vesicles.

\section{Model}
\label{model}

We consider a lipidic bilayer made of two types of lipids, A and B, where lipids A impose locally a spontaneous curvature and a larger thickness in a lipidic \textit{monolayer}. By symmetry, if the two monolayers have the same composition on the same site, the average spontaneous curvature due to lipid A is zero but the thickness can vary.
In the most general case, a global spontaneous curvature $C_0$ is imposed in equilibrium~\cite{helfrich1973}, in order to account for the bilayer nature of the membrane. This spontaneous curvature can be due to an asymmetry between the two leaflets, either in the number of lipids (or the leaflet areas), or in their lipidic composition. Note that this is the general case in biological cell membranes.

We will work at constant enclosed volume $V$, which defines an effective radius $R\equiv(\frac34 V/\pi)^{1/3}$ of the reference sphere corresponding to the sphere of volume $V$~\cite{seifert1995,seifert1997,barbetta,milner}. 
In the absence of thermal fluctuations (\textit{i.e.} formally at temperature $T=0$~K), the geometry of the membrane would then be a sphere of radius $R$. Due to thermal fluctuations at temperature $T>0$, the position of the membrane with respect to the center of the vesicle is given by~\cite{seifert1995,barbetta,milner}
\be
\br=r(\theta,\varphi)\hat {\bf e}_r=R[1+ u(\theta,\varphi)]\hat {\bf e}_r
\ee
where $(\hat {\bf e}_r,\hat {\bf e}_\theta,\hat {\bf e}_\varphi)$ is the natural base in spherical coordinates (see fig.~\ref{fig1}).

We consider the most general Hamiltonian of a lipidic bilayer depending on three fields: the dimensionless position field $u(\Omega)$ where we use for short the notations $\Omega\equiv(\theta,\varphi)$ and $d\Omega=\sin\theta d\theta d\phi$, and the surface fractions of the lipid of type A in the upper and lower layers, noted $\phi_1(\br)$ and $\phi_2(\br)$ respectively (the surface fraction of lipid of type B is therefore $1-\phi_i$ in the monolayer $i$). Formally, these fields are coarse-grained functions of the distributions $\hat\phi_i (\br)=s_A\sum_j\delta(\br-\br^{(i)}_j)$ where the lipids A of surface $s_A$ have their center of mass located at positions $\br^{(i)}_j$ on the leaflet $i$. We assume that for a planar monolayer, the average surface fraction $\bar \phi_i=N_i s_A/S_i$ are fixed, where $N_i$ is the average number of lipids A on the leaflet $i$ (and $S_i$ its area). In general, $N_1\neq N_2$.

The Hamiltonian is~:

\bea
 {\cal H}[\phi_1,\phi_2,u] = \int_\mathcal{A} \mathrm{d}\mathcal{A}  \left[\frac{J}2 g^{ij}\nabla_i\phi_1\nabla_j \phi_1 + \frac{m_1}2 (\phi_1 -\bar \phi_1)^2 \right. \nonumber\\
+\left.\frac{J}2 g^{ij}\nabla_i\phi_2\nabla_j \phi_2 + \frac{m_2}2 (\phi_2 -\bar \phi_2)^2 \right. \nonumber\\ 
+\left. \frac{k}2 [(\phi_1-\bar \phi_1)-(\phi_2-\bar \phi_2)]^2\right] \nonumber \\
+\sigma \mathcal{A} + \frac12\int_\mathcal{A} \mathrm{d}\mathcal{A}  \ \kappa(\phi_1+\phi_2)\left[\mathrm{div}(\bn) - C(\phi_1-\phi_2)\right]^2 \ \ \label{H0}  
\eea
where $\mathcal{A}$ is the bilayer area and $\sigma$ the bare surface tension. The first term is a Ginzburg-Landau hamiltonian for the lipid composition fields $\phi_1$ and $\phi_2$, living on a surface with metric $g_{ij}=\delta_{ij}+\nabla_i r \nabla_j r$ (the inverse metric tensor is $g^{ij} = \delta^{ij} - \nabla_i r \nabla_j r/g$ and $g = 1 + (\nabla r)^2$ is its determinant). We define the 2D gradients as $\nabla \equiv\frac1{r} \tnabla $ where $ \tnabla \equiv \hat {\bf e}_\theta\partial_\theta +\hat {\bf e}_\varphi\frac1{\sin\theta} \partial_\varphi$. The element of area is $\mathrm{d}\mathcal{A}=r^2\sqrt{g}\mathrm{d}\Omega$.

The positive energetic parameter $J$ favors lipids of the same type being next to each other, and the potentials $m_i(\phi_i -\bar \phi_i)^2/2$ ensures an homogeneous phase at high temperatures, $T>T_{\rm c}$, and a phase separation into two phases at lower temperatures, $T<T_{\rm c}$, where $T_{\rm c}$ is the critical temperature for a planar monolayer made of lipids A and B. The ``mass'' of the theory, $m_i$, which depends on $\bar\phi_i$, is the coefficient that defines the transition for a planar monolayer. Close to $T_{\rm c}$, it can be written as $m_i=a_i(T-T_{\rm c})$ in a Landau theory. In the following, we assume that $m_1$ and $m_2$ are positive, \textit{i.e.} that the homogeneous (or liquid) phase is stable for a planar and non-fluctuating monolayer. 
Note that higher order terms in $\phi^3$ or $\phi^4$ should be taken into account when the planar and non-fluctuating monolayer phase-separates, \textit{i.e.} when $m_i<0$. These terms are necessary, for instance, to compute the compositions in the two phases.

The quadratic term $(\phi_1-\bar{\phi}_1-\phi_2+\bar{\phi}_2)^2$ couples the composition of the two monolayers in a similar way as in the area-difference elasticity model where $k$ can be seen as a compression modulus~\cite{seifert1997}. It forbids large deviations in the composition of the two monolayers. Moreover if one takes a transversal interaction between both leaflets into account, the authors of~\cite{sunilkumar} show one obtains such a quadratic term.

The second part of \eq{H0} is a generalization of the Hamiltonian of the Helfrich spontaneous curvature model~\cite{helfrich1973} where the bilayer bending modulus $\kappa$ and the spontaneous curvature $C$ are functions of the sum $\phi_1+\phi_2$ and the difference $\phi_1-\phi_2$ of the lipid composition fields, respectively. 
The normal vector to the membrane is $\bn=(\hat {\bf e}_r-\nabla r)/\sqrt{g}$, and its divergence, $\mathrm{div} (\bn)$, is the local membrane curvature~\cite{helfrich1973,helfrich1986}.

We note $\phi_i= \bar \phi_i+\psi_i$ where we assume $\psi_i\ll\bar \phi_i$. Note that we do not require $\bar{\phi}_{\rm i}$ to minimize the mean-field energy, because we are generically interested in live cells that impose lipidic compositions of leaflets through active processes~\cite{simons1,lodish}. Alternatively, a freshly created vesicle needs a long time to reach equilibrium~\cite{ole}. Instead of imposing average values, one can use chemical potentials~\cite{leibler2,hirose} in the grand-canonical ensemble that control the leaflet compositions, which do not change the results (see Appendix~A). 

We choose a linear interpolation for the bending modulus and the spontaneous curvature:
\bea
\kappa &=& \kappa_0+\kappa_1\psi_+\\
C &=& C_0 +C_1 \psi_-
\eea
where we have introduced the fields,
\be
\psi_-=\frac{\psi_1-\psi_2}2;\qquad \psi_+=\frac{\psi_1+\psi_2}2
\ee
But contrary previous studies were $\phi_1+\phi_2$ is fixed and constant~\cite{safran1990,safran1991,mac1993}, we only fix it globally (that is to say its integral over the surface). Indeed, in ref.~\cite{sunilkumar}, the authors show that $\psi_+$ is "a thermodynamic variable, which can not be integrated out to give a simpler model". The mean bending modulus is $\kappa_0=\kappa_-+(\bar\phi_1+\bar\phi_2)\kappa_1/2$  and  the difference in bending rigidities $\kappa_1=\kappa_+-\kappa_-$ with $\kappa_-$ (respectively $\kappa_+$) the bending modulus of a membrane with lipids of type B (resp. A) only (see fig.~\ref{fig1}). The mean spontaneous curvature is $C_0= C_0^*+(\bar\phi_1-\bar\phi_2)C_1/2$ where $C_1$ is related to the lipid A geometry and $C_0^*$ is the bare spontaneous curvature of the pure B vesicle. In the case of a mono-component bilayer, it has been shown by Seifert~\cite{seifert1995} that the spontaneous curvature model of Helfrich leads to similar results as the area-difference elasticity model with $C^*_0=0$ (but with a constraint on the total mean curvature). However, in the present case, we will break the symmetry between the two leaflets through either the bare spontaneous curvature $C_0^*$ or the difference in compositions $\bar\phi_1-\bar\phi_2$.

Hence for regions where $\phi_1(\br)=\phi_2(\br)$, the spontaneous curvature vanishes and the bending modulus interpolates between $\kappa_-$ for $\phi_1(\br)=\phi_2(\br)=0$ and $\kappa_+>\kappa_-$ for $\phi_1(\br)=\phi_2(\br)=1$. Therefore, high values of $\phi_1$ and $\phi_2$ favors large bending rigidities and theses regions of the bilayer can be seen as thicker. It thus corresponds to the rich-cholesterol and sphingolipid phase~\cite{schmid,schick}. In addition, the local spontaneous curvature is due, in this model, to a mismatch between the compositions of the upper and the lower monolayer.

In the following, we will work at constant bare surface tension $\sigma$ for the purpose of modeling real cells. Indeed, in a cell membrane, the surface tension is on the order of $10^{-4}$~J/m$^2$, due to many factors, such as the membrane composition (lipids and proteins), and more specifically the presence of the cytoskeleton~\cite{dai}. Hence, contrary to many previous works on quasi-spherical vesicles (made of one type of lipids only)~\cite{seifert1995,barbetta,milner,helfrich1986}, we do not work at constant area, which, in the conventional approach, is forced by a Lagrange multiplier, the effective surface tension. The free parameters are thus the surface tension $\sigma$ and the two radii $R$ and $R_0\equiv2/C_0$, which enforce the volume and the mean spontaneous curvature.

The Hamiltonian~\eq{H0} becomes
\bea
{\cal H}= \sigma \mathcal{A} + \int_\mathcal{A} \mathrm{d}\mathcal{A} \left[J g^{ij}(\nabla_i\psi_+\nabla_j \psi_+ + \nabla_i\psi_-\nabla_j \psi_-) \right.\nonumber\\
\left.  + \frac{m_+}2\psi_+^2  + \frac{m_-}2\psi_-^2 +m_0\psi_+\psi_- \right] \nonumber \\ 
+\frac12\int_\mathcal{A} \mathrm{d}\mathcal{A} \,(\kappa_0+\kappa_1\psi_+)\left[\mathrm{div}(\bn) - C_0-C_1\psi_-\right]^2 
\label{H1}
\eea
where
\bea
m_+ &=& m_1+m_2 \label{mplus}\\
m_- &=& m_1+m_2+4k  \label{mmoins}\\
m_0 &=& m_1-m_2
\eea

In the following, we expand the full Hamiltonian, \eq{H1}, up to quadratic order, with terms in $u^2$, $u\psi$ and $\psi^2$. Hence, we limit ourselves to the so-called quasi-spherical vesicle approximation~\cite{seifert1995,barbetta}. Following Helfrich~\cite{helfrich1986}, we have~:
\bea
\mathrm{d}\mathcal{A}\equiv r^2\sqrt{g}\mathrm{d}\Omega \simeq R^2\left[1+2u+u^2+\frac12(\tnabla u)^2 \right]\,\mathrm{d}\Omega\\
\mathrm{div}(\bn)\, \mathrm{d}\mathcal{A} \simeq 2R\left[1+u-\frac12\tnabla^2u+\frac12(\tnabla u)^2 \right]\,\mathrm{d}\Omega\\
(\mathrm{div}\bn)^2\mathrm{d}\mathcal{A} \simeq 4 \left[1-\tnabla^2u+\frac14(\tnabla^2u)^2+ u\tnabla^2u\right.\nonumber\\
\left.+\frac12(\tnabla u)^2 \right]\,\mathrm{d}\Omega
\eea
and at this order the Landau term simplifies to 
\be
g^{ij}\nabla_i\phi \nabla_j \phi \mathrm{d}\mathcal{A} \simeq (\tnabla \phi)^2 \,\mathrm{d}\Omega
\ee
The Hamiltonian \eq{H1} can be separated in four parts~:
\bea
{\cal H}&=& E_{\rm sph}(R)+ {\cal H}_{\rm Helf}[u] +{\cal H}_{\rm GL}[\psi_+,\psi_-] \nonumber \\
&&+ \delta{\cal H}[\psi_-,\psi_+,u]
\label{H2}
\eea
where the first term
\be
E_{\rm sph}(R)=4\pi R^2 \sigma +2\pi\kappa_0(C_0R-2)^2
\ee
are the surface and bending energies of the reference sphere of radius $R$. The second term is the usual Helfrich Hamiltonian describing the height fluctuations
\bea
{\cal H}_{\rm Helf}[u] =  \frac{\kappa_0}2\int_\mathcal{S} \mathrm{d}\Omega \left[4(\tilde \sigma -c_0) u+ 2(c_0-2) \tnabla^2u \right.\nonumber\\ \left.+  2\tilde \sigma u^2+ (\tnabla^2 u)^2 + (\tilde\sigma+2c_0-2) (\tnabla u)^2 \right]
\label{Hh}
\eea
where $\tilde \sigma = \hat \sigma + c_0^2/2$.
For sake of clarity, we make all the parameters dimensionless by dividing all the lengths by the radius $R$ and the energies by the average bending modulus $\kappa_0$~:  $c_0=C_0R$, $c_1=C_1R$, $\hat J=J/\kappa_0$ and $\hat m_i= m_i R^2/\kappa_0$, and $\hat \sigma =\sigma R^2/\kappa_0=(R/\xi)^2$ where 
\be
\xi\equiv\sqrt{\frac{\kappa_0}{\sigma}}
\ee 
is the usual correlation length for planar membranes under tension.

The third term of \eq{H2} is the Ginzburg-Landau Hamiltonian written in the $(\psi_+,\psi_-)$ basis
\bea
{\cal H}_{\rm GL}[\psi_+,\psi_-] = \kappa_0 \int_\mathcal{S} \mathrm{d}\Omega \left[\hat J(\tnabla\psi_+)^2+\hat J(\tnabla\psi_-)^2\right.\nonumber\\\left.+ \tilde\mu_+\psi_+ + \tilde\mu_-\psi_- + \frac{\hat m_+}2\psi_+^2+ \frac{ \tilde m_-}2 \psi_-^2 + \tilde m_0 \psi_+\psi_-\right]
\label{GL}
\eea
with $\tilde \mu_+ = \hat \kappa_1 (2-c_0)^2$, $\tilde \mu_- =  - c_1(2-c_0)$, $\tilde m_- = \hat m_- + c_1^2$, and $\tilde m_0 = \hat m_0- \hat \kappa_1 c_1(2-c_0)$.

The last term of \eq{H2} is the coupling contribution:
\bea
\delta{\cal H}&=& \kappa_0 \int_\mathcal{S} \mathrm{d}\Omega \left\{\left[c_1\psi_--\hat \kappa_1(2-c_0)\psi_+\right]\tnabla^2u \right.\label{deltaH}\\ 
&+&\left. 2 c_1(c_0-1)u\psi_- -\hat\kappa_1c_0(2-c_0)u\psi_+ \right\}\nonumber
\eea

To proceed further, we follow the lines of ref.~\cite{milner} and we decompose the Hamiltonian \eq{Hh} using the standard decomposition of $u(\theta,\varphi)$ in spherical harmonics
\be
u(\theta,\varphi)=\frac{u_{00}}{\sqrt{4\pi}}+\sum_\lambda u_{lm}Y_l^m(\theta,\varphi)
\label{harmonics}
\ee
where $\sum_\lambda=\sum_{l=1}^{l_{\max}}\sum_{m=-l}^l$ with $l_{\max}\simeq R\Lambda$ being an ultraviolet cutoff (where $\Lambda^{-1}\simeq1$ to 5~nm is on the order of the membrane thickness). Furthermore, following~\cite{seifert1995,milner,safran}, we impose a constant volume to the cell, which supposes that the membrane is impermeable to water. We thus find
\be
V\equiv\frac{4\pi}3R^3=\frac{R^3}3\int_\mathcal{S} \mathrm{d}\Omega[1+u(\Omega)]^3
\ee
which implies
\be
u_{00}\sqrt{4\pi} =-\sum_\lambda |u_{lm}|^2
\ee
It is thus straightforward to show that the mode $l=1$ does not modify the cell area
\be
\mathcal{A}=4\pi R^2+R^2\sum_\lambda\left[\frac{l(l+1)}2-1\right]|u_{lm}|^2\label{area}
\ee

By injecting this equality in the decomposition of \eq{Hh} and using the usual properties of the spherical harmonics, we obtain
\be
{\cal H}_{\rm Helf} = \frac{\kappa_0}2 \sum_\lambda H(l) |u_{lm}|^2
\ee
where
\be
H(l)=[l(l+1)-2]\left[l(l+1)+\hat \sigma - c_0\left(2-\frac{c_0}2\right)\right]
\label{Hl}
\ee
a result which has been previously obtained by Milner and Safran~\cite{milner}. We have $H(l=1)=0$ which is the signature that the three modes $l=1$ correspond to the simple translation of the vesicle in the three directions of space without cost in bending and surface energies. Hence in the following the sum $\sum_\lambda$ will be restricted to the values $l\geq2$.

By decomposing in spherical harmonics the lipid composition fields following \eq{harmonics} 
\bea
\psi_- &=& \frac{a_{00}}{\sqrt{4\pi}} + \sum_\lambda a_{lm}Y_l^m(\theta,\varphi)\\
\psi_+ &=& \frac{b_{00}}{\sqrt{4\pi}} + \sum_\lambda b_{lm}Y_l^m(\theta,\varphi)
\eea
we find for \eq{GL}~\footnote{Due to the property $a^*_{lm}=(-1)^m a_{l-m}$, one has $\sum_{m=-l}^l(b_{lm}a^*_{lm}+b^*_{lm}a_{lm})=2\sum_{m=-l}^lb_{lm}a^*_{lm}=2\sum_{m=-l}^lb^*_{lm}a_{lm}$.}
\bea
{\cal H}_{\rm GL} &=& \kappa_0(2\sqrt{4\pi} (\tilde \mu_+ b_{00} +\tilde \mu_- a_{00}) +  \tilde m_0 b_{00}a_{00}) \nonumber\\ 
&+& \kappa_0 \sum_\lambda \left\{ \left[\frac{\hat m_+}2+ \hat J l(l+1)\right] |b_{lm}|^2\right.\nonumber\\
 &+&\left.\left[ \frac{ \tilde m_-}2 + \hat J l(l+1)\right] |a_{lm}|^2+ \tilde m_0 b_{lm}a^*_{lm}\right\}
\eea
and the coupling Hamiltonian \eq{deltaH}
\be
\delta{\cal H} = \kappa_0 \sum_\lambda [\Theta_+(l)b_{lm}^*+ \Theta_-(l)a_{lm}^*] u_{lm} 
\ee
where
\bea
\Theta_+(l) &=& \hat\kappa_1(2-c_0)[l(l+1)-c_0] \label{thetaplus}\\
\Theta_-(l) &=& -c_1[l(l+1)+2-2c_0] \label{thetamoins}
\eea

\section{Correlation functions in lipid composition}

The full Hamiltonian being quadratic in $u_{lm}$, we partially integrate it following
\bea
\mathcal{Z} &\equiv& e^{-\beta E_{\rm sph}} \int \mathcal{D} \psi_- \mathcal{D} \psi_+ \,e^{-\beta {\cal H}_{\rm GL}}\int \mathcal{D}u \,e^{-\beta ({\cal H}_{\rm Helf}+\delta{\cal H})} \nonumber\\
&=& e^{-\beta E_{\rm sph}} \sqrt{\frac{2\pi}{\det (\beta \mathcal{H}_{\rm Helf})}}\int \mathcal{D} \psi_- \mathcal{D} \psi_+ \,e^{-\beta {\cal H}_{\rm eff}}
\eea
where $\mathcal{D}u=\prod_{l=2}^{l_{\max}}\left(\prod_{m=0}^l \mathrm{d}\mathrm{Re}(u_{lm})\prod_{m=1}^l \mathrm{d}\mathrm{Im}(u_{lm})\right)$~\cite{seifert1995}. The effective Ginzburg-Landau Hamiltonian thus becomes
\be
{\cal H}_{\rm eff} = \frac{\kappa_0}2\sum_\lambda \bB_{lm}^\dag \bM(l) \bB_{lm}
\label{Heff}
\ee
where $\bB_{lm}^\dag=(b_{lm}^*,a_{lm}^*)$ and the three elements of the symmetric matrix $\bM(l)$ are
\bea
M_+(l) &=& \hat m_+ + 2\hat J l (l+1) - \frac{\Theta_+^2(l)}{H(l)} \label{Mplus}\\
M_-(l) &=& \hat m_- + c_1^2 + 2\hat J l (l+1) - \frac{\Theta_-^2(l)}{H(l)}\label{Mmoins}\\
M_0(l) &=& \hat m_0 - \hat \kappa_1 c_1(2-c_0) -\frac{\Theta_+(l)\Theta_-(l)}{H(l)} \label{M0}
\eea
The three static structure factors $S_+(l)$, $S_-(l)$, and $S_0(l)$ are thus the elements of $\bM^{-1}(l)$.  As \eq{Mmoins} shows it, the transversal coupling  constant $k$ plays a role through $\hat{m}_{-}$: we shall see below that the mass of the system increases such that the divergence of $S_-(l)$ occurs for larger values of $\hat{m}_0$. The total free energy of the membrane is
\bea
\mathcal{F}&=& E_{\rm sph}(R) + \frac{k_{\rm B}T}2 \sum_{l=2}^{l_{\max}}(2l+1)\left\{\ln[\beta\kappa_0 H(l)]\right. \nonumber\\&&+\left.\ln[(\beta\kappa_0)^2 (M_+(l)M_-(l)-M_0^2(l))]\right\}
\eea
The expectation value of the area, given by $\langle \mathcal{A} \rangle=\frac{\partial\mathcal{F}}{\partial \sigma}$, is
\bea
\frac{\langle \mathcal{A} \rangle}{4\pi R^2 } &=&  1+ \frac{k_{\rm B}T}{8\pi\kappa_0}\sum_{l=2}^{l_{\max}}\frac{2l+1}{l(l+1)+\hat \sigma - c_0\left(2-\frac{c_0}2\right)}\nonumber\\
&\times&\left[1+ \frac{M_+\Theta_-^2+M_-\Theta_+^2-2M_0\Theta_-\Theta_+}{H(M_+M_--M_0^2)}\right] 
\eea
In the bracket, the first term equal to 1 is the usual excess area for a mono-component bilayer~\cite{seifert1995,helfrich1986}, and the second term corresponds to the area induced by the lipidic composition of the two leaflets. Note that for mono-component bilayers, the apparent surface tension $\hat\sigma'=\hat \sigma-c_0(2-c_0/2)$, which is renormalized by $c_0$, can be negative. However, it does not correspond to the mechanical surface tension $\tau$, which can be even more negative. For instance $\hat\sigma'=-4$, \textit{i.e.} $\sigma\simeq10^{-8}$~N/m (for $R=5~\mu$m and $\kappa_0=10^{-19}$~J), which is the lower limit for quasi-spherical membranes, yields an excess area of 2.5\% and a mechanical tension $\tau\simeq-10^{-5}$~N/m)~\cite{barbetta}. Very large values of $\hat\sigma'> l_{\max}^2$ leads to very small excess area and measurable mechanical surface tensions of $\tau\simeq10^{-4}$~J/m$^{-2}$. 

The correlation functions of the fields $\psi_-$ and $\psi_+$ in two points $\br_1$ and $\br_2$ are, due to the isotropy on the sphere, functions of the angle $\gamma$ between the two vectors $\br_1$ and $\br_2$, given by
\be
\cos\gamma = \cos\theta_1\cos\theta_2+\sin\theta_1\sin\theta_2\cos(\phi_1-\phi_2)
\ee 
Hence we have
\bea
\langle \psi_-(\gamma)\psi_-(0)\rangle = \frac{k_{\rm B}T}{4\pi} \sum_{l\geq2} \frac{(2l+1)P_l(\cos\gamma)}{M_-(l)-M_0^2(l)/M_+(l)} \label{AngCorrmoins}\\
\langle \psi_+(\gamma)\psi_+(0)\rangle = \frac{k_{\rm B}T}{4\pi} \sum_{l\geq2} \frac{(2l+1)P_l(\cos\gamma)}{M_+(l)-M_0^2(l)/M_-(l)}\label{AngCorrplus}\\
\langle \psi_-(\gamma)\psi_+(0)\rangle = \frac{k_{\rm B}T}{4\pi} \sum_{l\geq2}\frac{(2l+1)M_0(l)P_l(\cos\gamma)}{M_0^2(l)-M_+(l)M_-(l)}\label{AngCorrpm}
\eea
where $P_l(x)$ are the Legendre polynomials. These are complicated expressions with 9 independent parameters: $c_0$, $c_1$, $\hat \kappa_1$, $\hat \sigma$, $\hat J$, $\hat m_\pm$, $\hat m_0$, $\hat \mu_-$ which are related to the temperature $T$, the average compositions in each monolayers initial  $\bar\phi_1,\bar\phi_2$, and the initial parameters $k,C_0^*,C_1,\kappa_-,\kappa_+,J,\sigma$, and $R$. In the next Section, we first consider the simplest case where the membrane is planar on average, \textit{i.e.} $R\to\infty$.

\section{Planar membrane}

\subsection{Planar case as the limit $R\to\infty$}

Here we want to compare our results to the previous works on the curvature-mediated model for planar membranes\cite{leibler1,leibler2,mackintosh,schick,hirose}. By choosing a squared patch of area $A=4\pi R^2$, the wave-vectors are quantified following
$\bq=2\pi/\sqrt{A} (n_x,n_y)$~\cite{barbetta}. The number of modes in a  corona of radius $q$ and thickness $\mathrm{d}q$ is thus $2\pi q \mathrm{d}q/(2\pi/\sqrt{A})^2$. This should be equal in the limit $R\to\infty$ to $(2l+1)\Delta l$ (where $\Delta l=1$). Taking the  continuous limit, we find $Aq\mathrm{d}q/(2\pi)=2R^2q\mathrm{d}q=(2l+1) \mathrm{d}l$ and therefore $q^2=l(l+1)/R^2$. Moreover, previous works were restricted to the case $C_0=0$. Here we need a prescription to reach this limit when $R\to\infty$, which enforces the choice $c_0\to2$ (see Appendix~B for details).
We thus find $H(l)\to H(q)= R^4 q^2(q^2+ \xi^{-2})$ where $q\geq0$. Likewise, using \eqs{thetaplus}{thetamoins}, and keeping the terms of order $R^2$, we have
\bea
\frac{\kappa_0 M_-}{R^2} &\to& M_-(q)= m_- + 2Jq^2+ \frac{\kappa_0 C_1^2}{1+(q\xi)^2}\\
\frac{\kappa_0 M_+}{R^2} &\to& M_+(q)=m_++2Jq^2\\
\frac{\kappa_0 M_0}{R^2} &\to& M_0(q)=m_0
\eea 
Hence, by writing $b_{lm}\to \psi_+(\bq)/(4\pi R^2)$, in the limit $R\to\infty$, the Hamiltonian, \eq{Heff}, simplifies to 
\bea
{\cal H}_{\rm eff} &=& \frac12\int_0^\infty \frac{q \mathrm{d}q}{2\pi}\left\{ M_+(q)|\hat\psi_+(\bq)|^2  + M_-(q)|\hat\psi_-(\bq)|^2  \right. \nonumber\\ 
&+& \left. m_0 [\hat\psi_+(\bq) \hat\psi^*_-(\bq)+\hat\psi^*_+(\bq) \hat\psi_-(\bq)]\right\}
\label{HeffQ}
\eea
This is the expected Hamiltonian in the Fourier space for a planar membrane where the local curvature varies with the local composition $C(\psi_-)=C_1\psi_-$~\cite{leibler2,schick,hirose}. Note that for planar membranes the dependence of $\kappa(\psi_+)$ does not play any role at the Gaussian level. This comes from the fact that for a planar membrane $C_0=0$, \textit{i.e.} there is no constant spontaneous curvature. Note that an expansion in cumulants beyond the Gaussian level makes the effect of $\kappa_1$ observable~\cite{dean}. This is a major difference with the spherical case. The free energy of the planar membrane of area $\mathcal{S}$ becomes
\be
\mathcal{F}_{\rm plane}= \sigma \mathcal{S}  + \frac{k_{\rm B}T}2 \int_0^\infty \frac{q \mathrm{d}q}{2\pi} \ln\left[\frac{\beta^3\kappa_0q^2(q^2+\xi^{-2})}{S_+(q)S_-(q)-S_0^2(q)}\right]
\ee
where the inverses of the structure factors $S_{\pm,0}(q)$ are 
\bea
S^{-1}_-(q) &=& 2J q^2+ m_- + \frac{\kappa_0C_1^2}{1+(\xi q)^2}-\frac{m_0^2}{m_+ + 2Jq^2} \label{Smoins}\\
S^{-1}_+(q) &=& 2J q^2+ m_+ - \frac{m_0^2}{m_- + 2Jq^2 +\frac{\kappa_0C_1^2}{1+(\xi q)^2}} \label{Splus}\\
S^{-1}_0(q) &=& m_0 - \left(2J q^2+ m_- + \frac{\kappa_0C_1^2}{1+(\xi q)^2}\right)\frac{2J q^2+ m_+}{m_0}\nonumber\\
\eea

\subsection{Phase diagram}

To begin with, one notices that, for $m_0=0$, $S_-$ has a maximum at a non-zero wave-vector defined by
\be
(\xi q^*)^2=\frac{C_1}{C_1^*}-1\quad\mathrm{for}\quad C_1>C_1^*\equiv\frac{\sqrt{2J\sigma}}{\kappa_0}
\label{C1star}
\ee
As compared to refs.~\cite{sunilkumar,hirose}, we obtain n different value for $q^*$. Indeed, using our notations, they obtain~: $(q^* \xi)^2 \propto 1-(C_1^*/C_1)^2$.  Bot expressions are equivalent at small $C_1/C_1^*$. The main difference is that they obtain a saturation for large $C_1/C_1^*$, that we do not get. Following Hirose \textit{et al.}~\cite{hirose} we call this region Structured Disordered (SD) phase although it is not a true new phase in the thermodynamic sense but a liquid phase with high correlations in lipid composition associated to membrane height fluctuations between curved patches, with a correlation length on the order of $\xi$.  The wave-vector $q^*$ ($\propto (\sigma/2J)^{1/4}\sqrt{C_1}$ for large $C_1$ values) is thus the result of a balance between the spontaneous curvature $C_1$ which favors a large number of curved patches, and thus a small separation between them ($\simeq 1/q^*$), and the line tension, characterized by $J$, which tend to decrease the number of curved patches (and thus decrease $q^*$) which leads to a shorter total perimeter.

However the formation of mesophases (related to the divergence of $S_-$ at $q^*$) is not possible for $m_0=0$, since $S_-$ remains always positive as it can be checked from \eq{Smoins}. This is in contradiction with previous results~\cite{leibler1,hirose,schick,sunilkumar,shlomovitz}, where the coupling is introduced by a term in $-\kappa_0C_1\int \psi_-\nabla^2 h $ contrary to the natural hamiltonian $\frac12 \kappa_0\int (\nabla^2 h -C_1\psi_-)^2$. The last one introduces a renormalization of the mass by $\kappa_0C^2_1$ in \eq{HeffQ}, and forbids the divergence of the structure factor at $q^*$.
\begin{figure}[!t]
\centering
\includegraphics[width=.45\textwidth]{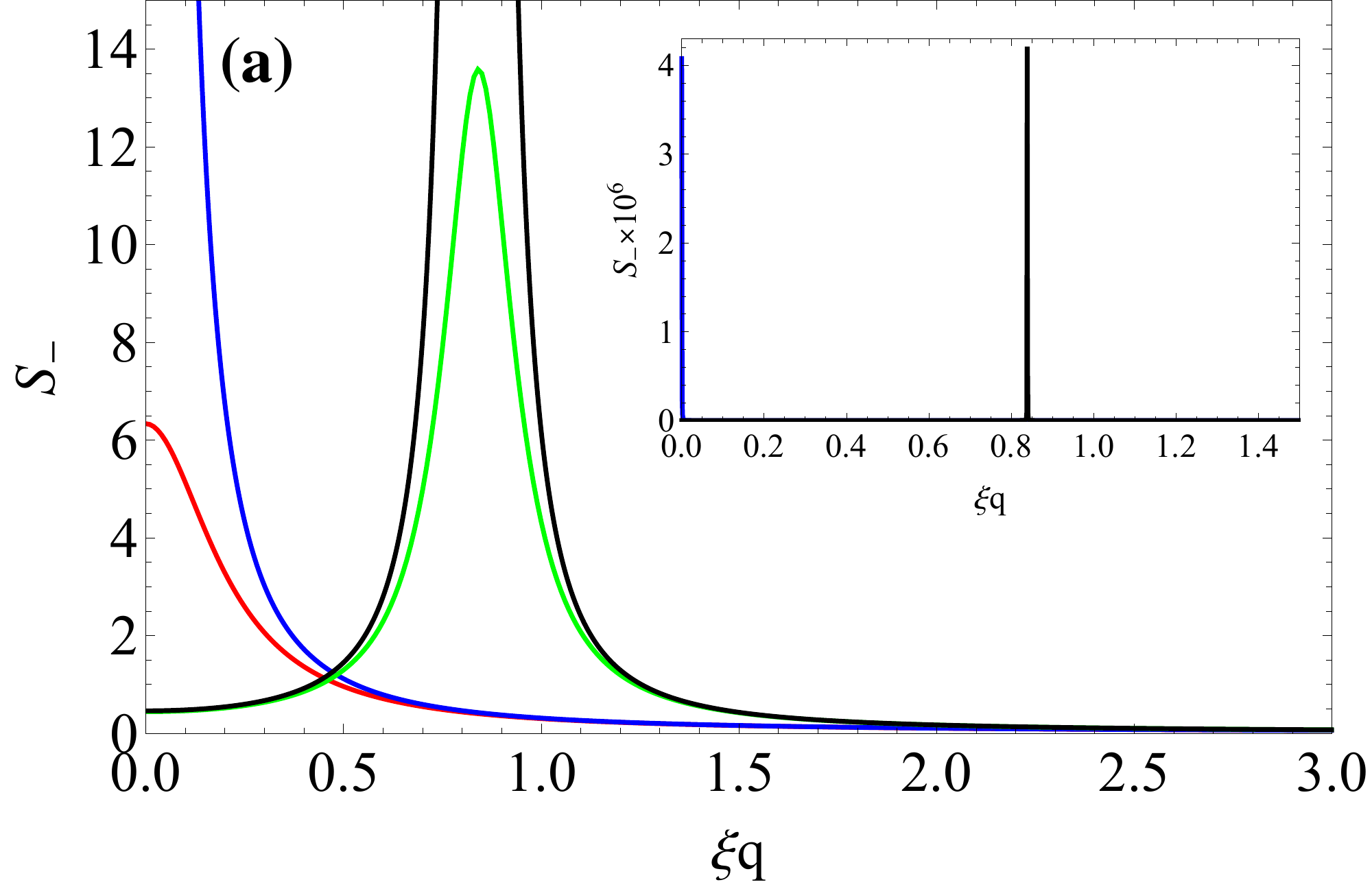}
\includegraphics[width=.45\textwidth]{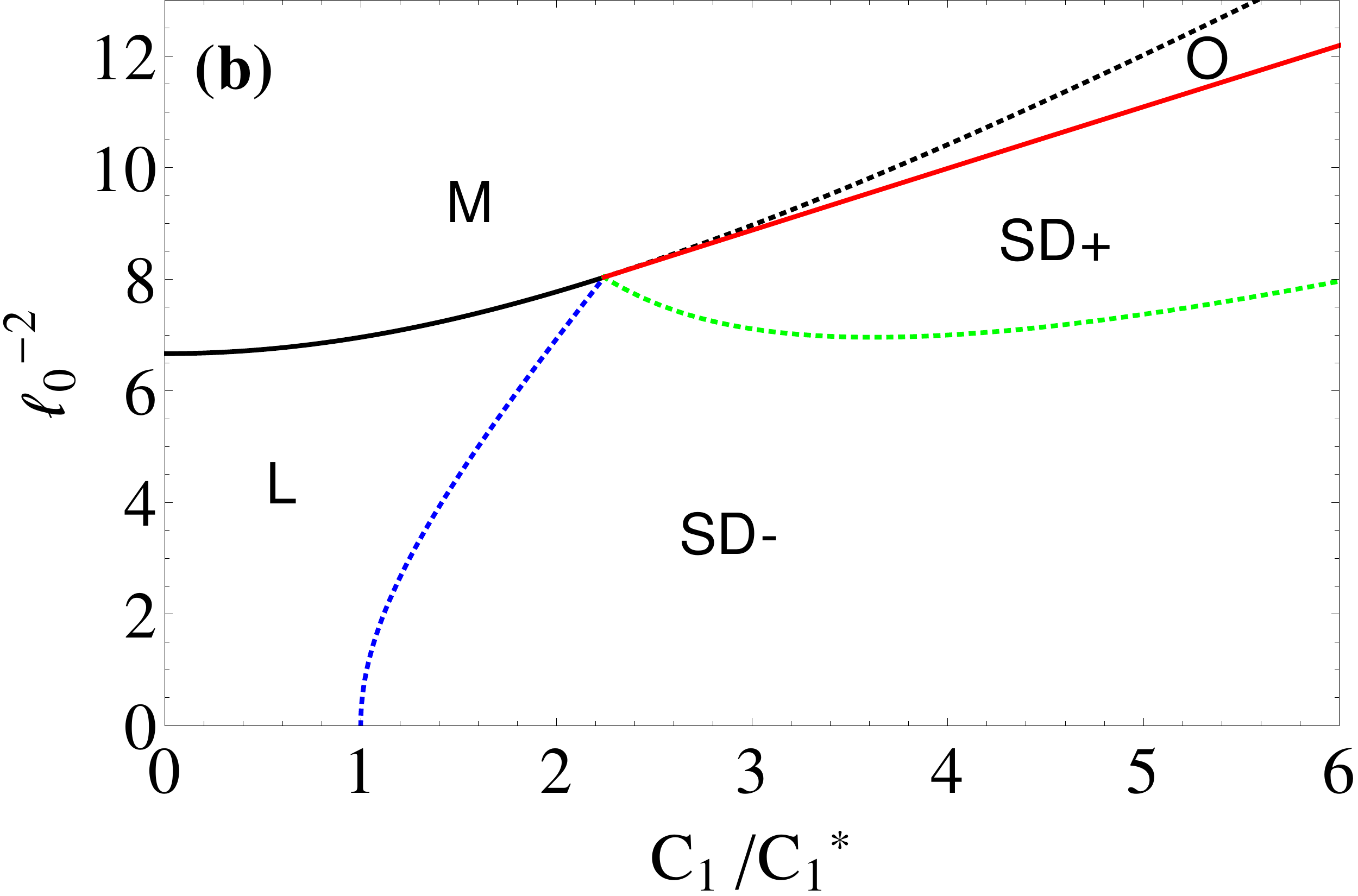}
\includegraphics[width=.45\textwidth]{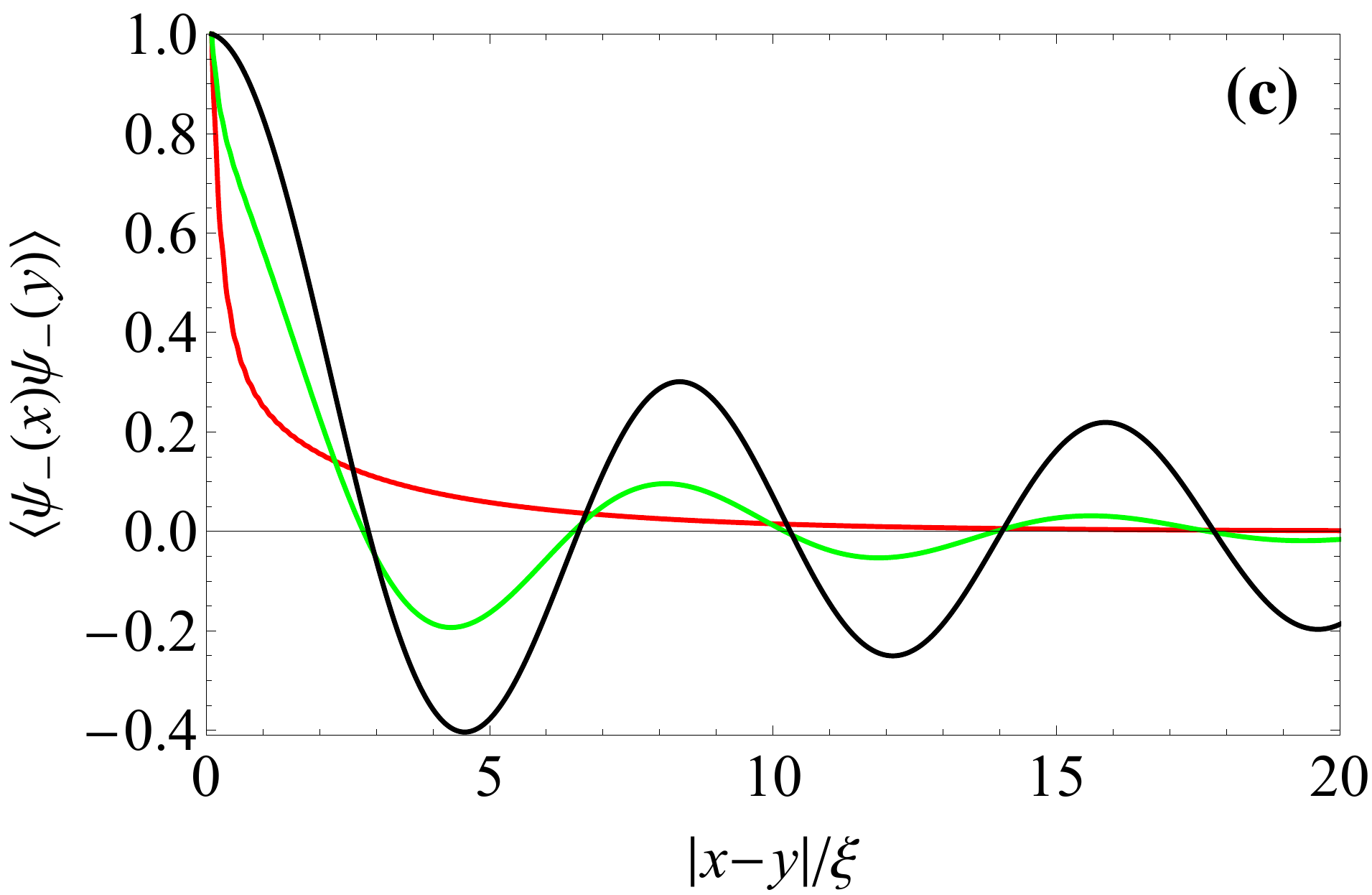}
\caption{\label{fig2} \textbf{(a)}~Dimensionless structure factor $\tilde S (\tilde q)$ corresponding to the 4 regions of the phase diagram~(b)  ($\ell_+/\xi =0.5$, $\ell_-/\xi =0.3$):  Liquid phase (L) in red ($(\xi/\ell_0)^2=6.64$, $C_1/C_1^*=0.25$); Macrophase separation (M) in blue ($(\xi/\ell_0)^2=6.69$, $C_1/C_1^*=0.25$); Structure Disordered (SD-) in green ($(\xi/\ell_0)^2=9.97$, $C_1/C_1^*=4$); and Ordered phase (O) in black ($(\xi/\ell_0)^2=9.99$, $C_1/C_1^*=4$). Inset: Blue and black curves on a larger scale. \textbf{(b)}~Corresponding phase diagram where $(\xi/\ell_0)^2\propto m_0$ is the coupling between $\psi_+$ and $\psi_-$, and $C_1$ is the local curvature induced by $\psi_-$. \textbf{(c)}~Correlation function associated to the three structure factors shown in (a)~: L (in red), SD- in green and O in black.}
\end{figure}

For a non-zero coupling constant $m_0$, \textit{i.e.} for different compositions in the two leaflets, divergence at finite wave-vector can occur leading to the formation of mesophases. Figure~\ref{fig2}(a) shows the adimensional structure factor $\tilde S(\tilde q)=2J S(q)/\xi^2$ where $\tilde q =\xi q$ for $\ell_+\equiv \sqrt{2J/m_+}=0.5\,\xi$ and $\ell_-\equiv \sqrt{2J/m_-}=0.3\,\xi$ (note that from \eqs{mplus}{mmoins}, we must have $\ell_+>\ell_-$) and for various values of $\ell_0^{-2}\equiv m_0/2J$ and $C_1/C_1^*$. We classify the behaviour of $\tilde S_-(\tilde q)$ in 4 cases which corresponds to the 4 phases shown in fig.~\ref{fig2}(b):\\
\textit{(i)}~For small values of $m_0$ and $C_1\lesssim C_1^*$ [red curve in fig.~\ref{fig2}(a)], $\tilde S_-(\tilde q)$ has a maximum at $\tilde q=0$ and decreases monotonously when $\tilde q$ increases. This behaviour  corresponds to the Liquid (L) phase delimited by the blue dotted and black lines in fig.~\ref{fig2}(b). The black line corresponds to the occurrence of a macrophase separation, \textit{i.e.} for $S^{-1}(0)=0$ which yields 
\be
\ell_0^{-2}=\ell_+^{-1}\sqrt{\ell_-^{-2}+\xi^{-2}(C_1/C_1^*)^2} \qquad (\mathrm{L/M})
\ee
\textit{(ii)}~For higher values of $\ell_0^{-2}$, the system follows a macrophase (M) separation and $S_-(q)$ exhibits a divergence at 0 [blue curve in fig.~\ref{fig2}(a)].\\
\textit{(iii)}~For $C_1>C_1^*$  and small $m_0$ values [green curve in fig.~\ref{fig2}(a)], $S_-(q)$ has a maximum at $q^*\neq0$. The SD- region is defined by the existence of $q_-^*\neq 0$ such that $S_-'(q_-^*)=0$. Similarly the Structure Disordered phase for thick patches (SD+) is defined by the existence of $q_+^*\neq 0$ such that $S_+'(q_+^*)=0$. The equations of the blue and green dotted lines are:
\bea
\ell_0^{-2} &=& \ell_+^{-2}\sqrt{(C_1/C_1^*)^2-1}  \qquad (\mathrm{L/SD-})\\
\ell_0^{-2} &=& \frac{\ell_-^{-2}+\xi^{-2}(C_1/C_1^*)^2}{\sqrt{(C_1/C_1^*)^2-1}}  \qquad (\mathrm{SD-/SD+})
\label{LSD+}
\eea
We have checked numerically that for $m_0>0$, $(\xi q^*)^2 \propto \frac{C_1}{C_1^*}\sqrt{1+(\ell_+/\ell_0)^4}-1$. Hence, as shown in  fig.~\ref{fig2}(b), there is a whole set of parameter values for which thick and thin patches are not correlated whereas curved ones are highly correlated (SD-).\\
\textit{(iv)}~Finally, for large $C_1/C_1^*$  and intermediate $m_0$ values [black curve in fig.~\ref{fig2}(a)], there is a small region in the phase diagram where both $S_-(q)$ and $S_+(q)$ diverges at $q^*\neq0$, which corresponds to the occurrence of mesophases (Ordered phase). Clusters of different lipidic composition are formed. The red line in fig.~\ref{fig2}(b) is defined through $\det M (q^2) \equiv M_-(q^2)M_+(q^2)-m_0^2=0$ and $\mathrm{d}(\det M)/\mathrm{d}(q^2)=0$.

The four separating lines meet at the tricritical point, the coordinates of which are 
\bea
\frac{C_{\rm 1, T}}{C_1^*} &=& \sqrt{\frac{1+(\ell_+/\ell_-)^2}{1-(\ell_+/\xi)^2}} \\
\ell_{\rm 0,T}^{-2} &=& \ell_+^{-1}\ell_-^{-1}\sqrt{\frac{\xi^2+\ell_-^2}{\xi^2-\ell_+^2}}
\eea
For $\ell_+=\xi$, the tricritical point escapes to infinity and no mesophase can be observed. Moreover, the line L/SD+ goes in the M phase and there are no SD phase for thick and thin lipid patches. Note that we focus on the SD phases since our model does not have any fourth order term. But with the asymmetry parameter $m_0$, it is possible to have an ordered phase only if we already have a SD one. Furthermore in~\cite{sunilkumar,taniguchi1994}, the authors study the effect of the temperature $T$ on the phase diagram. In our case, that would correspond to a change in the masses $m_1$ and $m_2$ that we do not consider into detail because we are mainly interested in the role of membrane physical parameters at a given temperature.

The correlation functions are computed by inverse Fourier transform 
\be
\langle \psi_-(r)\psi_-(0) \rangle = \int_0^\infty  \frac{q\mathrm{d}q}{2\pi} J_0(qr)S_-(q) 
\ee
Three examples of the normalized correlation functions are shown in fig.~\ref{fig2}(c) for the L, SD- and O phase. The L phase (red curve) exhibits a single correlation length on the order of $\ell_-$ whereas a structuration emerges in the SD-, shown by oscillations of period $2\pi/q*$, the maximum amplitude of which is obtained in the ordered phase. In this phase, a single mode is selected and the correlation function simplifies to $\langle \psi_-(r)\psi_-(0)\rangle= \frac{q^*}{2\pi}J_0(q^*r)$.

\section{Spherical membranes}

For spherical membranes, the spontaneous curvature $C_0$ and the bending modulus $\kappa_1$ enter into play.  Apart in the renormalized surface tension, $C_0$ also appears in the functions $\Theta_\pm(l)$ in \eqs{thetaplus}{thetamoins}  and in $M_0(l)$ [\eq{M0}] together with the parameters $\hat \kappa_1$ and $c_1$, which will impact the lipidic phases and the coupling between thick and curved patches.

Due to the large set of independent parameters, we only show some examples of the angular correlation functions of the lipidic fields $\psi_-$ and $\psi_+$ and discuss qualitatively the influence of these three parameters $c_0$, $c_1$ and $\hat \kappa_1$.
We first study the case of $\kappa_1=0$ to compare to the planar case and then look at the role played by $\kappa_1$. 

\subsection{Curvature-induced mechanism only ($\kappa_1=0$)}
\begin{figure}[t]
\includegraphics[width=.45\textwidth]{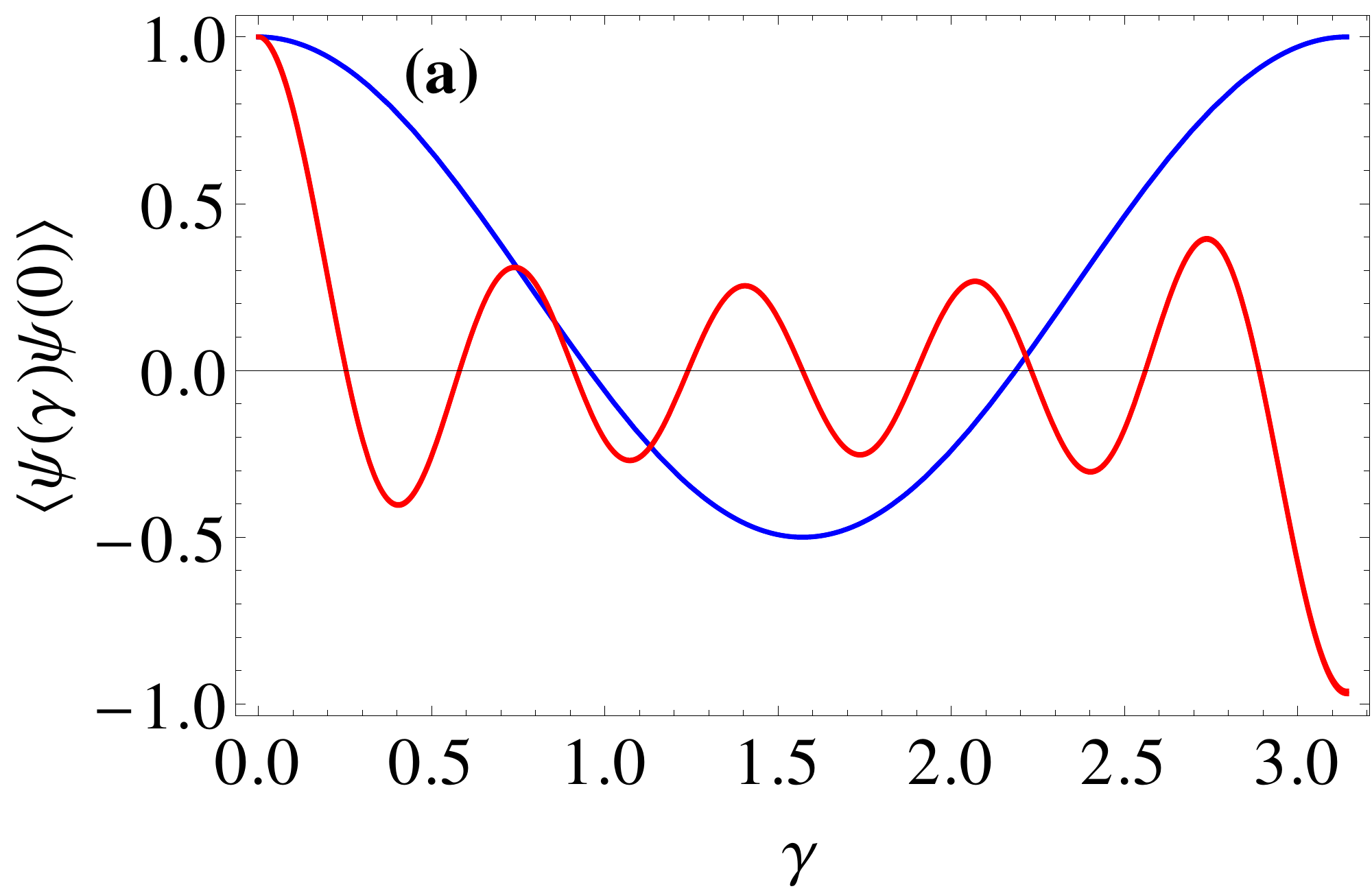}
\includegraphics[width=.44\textwidth]{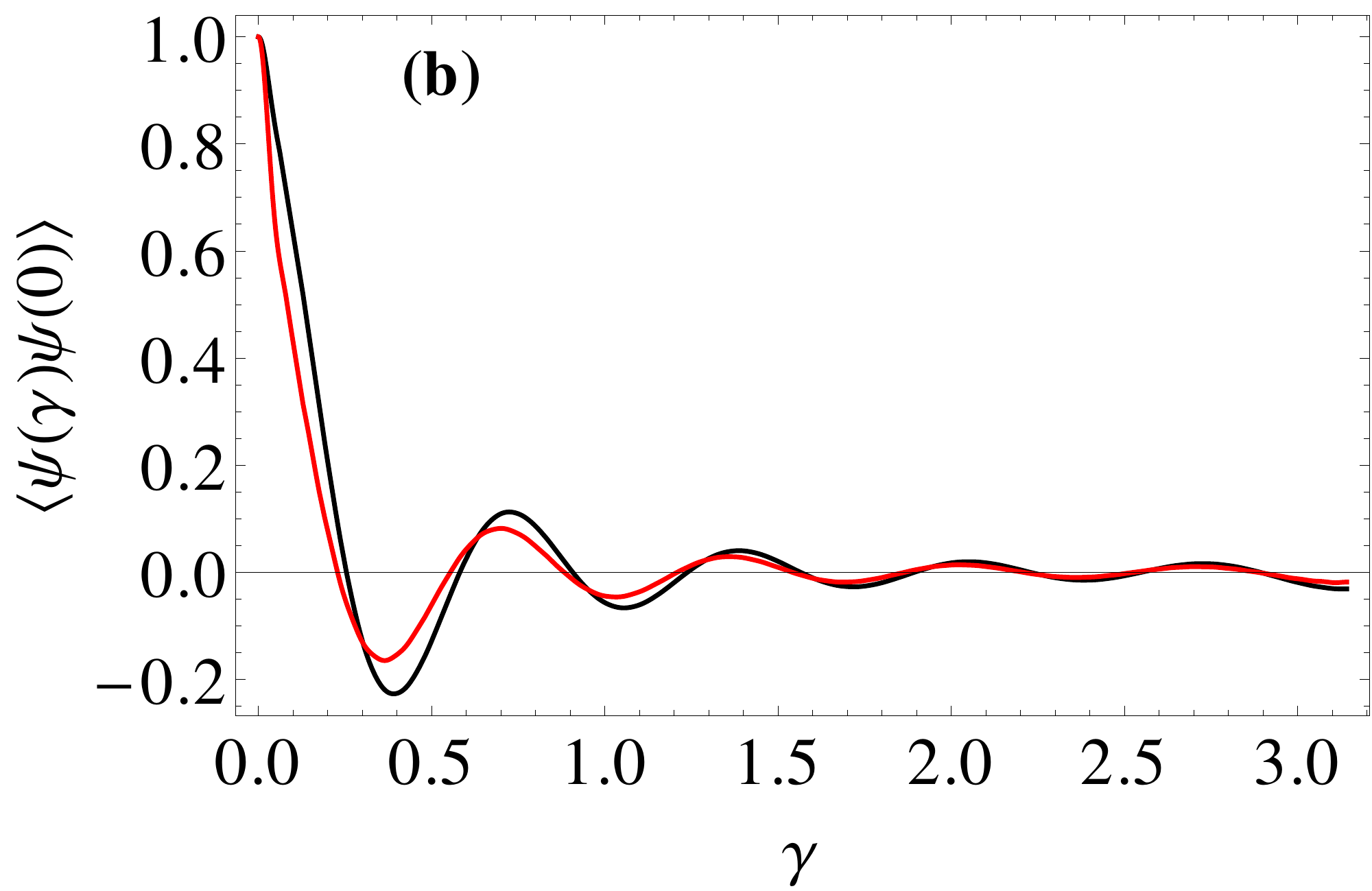}
\caption{\label{fig3} Normalized angular correlation functions, \eqs{AngCorrmoins}{AngCorrplus}, for a spherical membrane with $\kappa_1=0$ ($\hat m_+=\hat m_-=10$, $c_0=1.8$, $\hat J=0.05$). \textbf{(a)} In the L phase close to the macrophase transition (blue) ($\hat \sigma= 1$, $c_1=1.8$, $\hat m_0=1.03$) and close to the O one (red) ($\hat \sigma= 25$, $c_1=11.6$, $\hat m_0=1.95$). Both functions on $\psi_-$ and $\psi_+$ are superimposed. \textbf{(b)} In the SD phase for $\psi_+$ (black) and $\psi_-$ (red) ($\hat \sigma=25$, $c_1=11.6$, $\hat m_0=1.94$).}
\end{figure}
First, due to the finite size of the system, no true phase transition can occur and the crossover between the different phases is smooth. This is reflected by the fact that the divergence of the structure factor never occurs exactly for an integer value of $l$. Figure~\ref{fig3}(a) shows the correlation functions given by \eqs{AngCorrmoins}{AngCorrpm} in the liquid phase, close to the ``macrophase'' separation, and close to the ``ordered'' phase. First of all, one finds a very similar behavior as in the planar case. However, since the larger mode is for $l=2$, the correlation function has the C$_\infty$ symmetry and the system phase separates in two phases, the first one on the two poles $\gamma=0$ and $\pi$ (maxima of the correlation function) and the second around the equator $\gamma=\pi/2$ (blue curve). In the O phase, these two poles are anti-correlated (red curve) and, for these parameter values, 5 maxima appear. Note that, as in the planar case, $\langle \psi_-(\gamma)\psi_-(0)\rangle$ and $\langle \psi_+(\gamma)\psi_+(0)\rangle$ are identical since the divergence of the structure factor comes from the fact that the determinant [the denominator of \eqs{AngCorrmoins}{AngCorrpm}] reaches zero. The typical correlation functions in the SD phase are shown in fig.~\ref{fig3}(b), again showing a fast damping (on $\approx1$~rad for this set of parameter values).

\subsection{Bending-induced mechanism ($\kappa_1>0$)}
\begin{figure}[t]
\includegraphics[width=.45\textwidth]{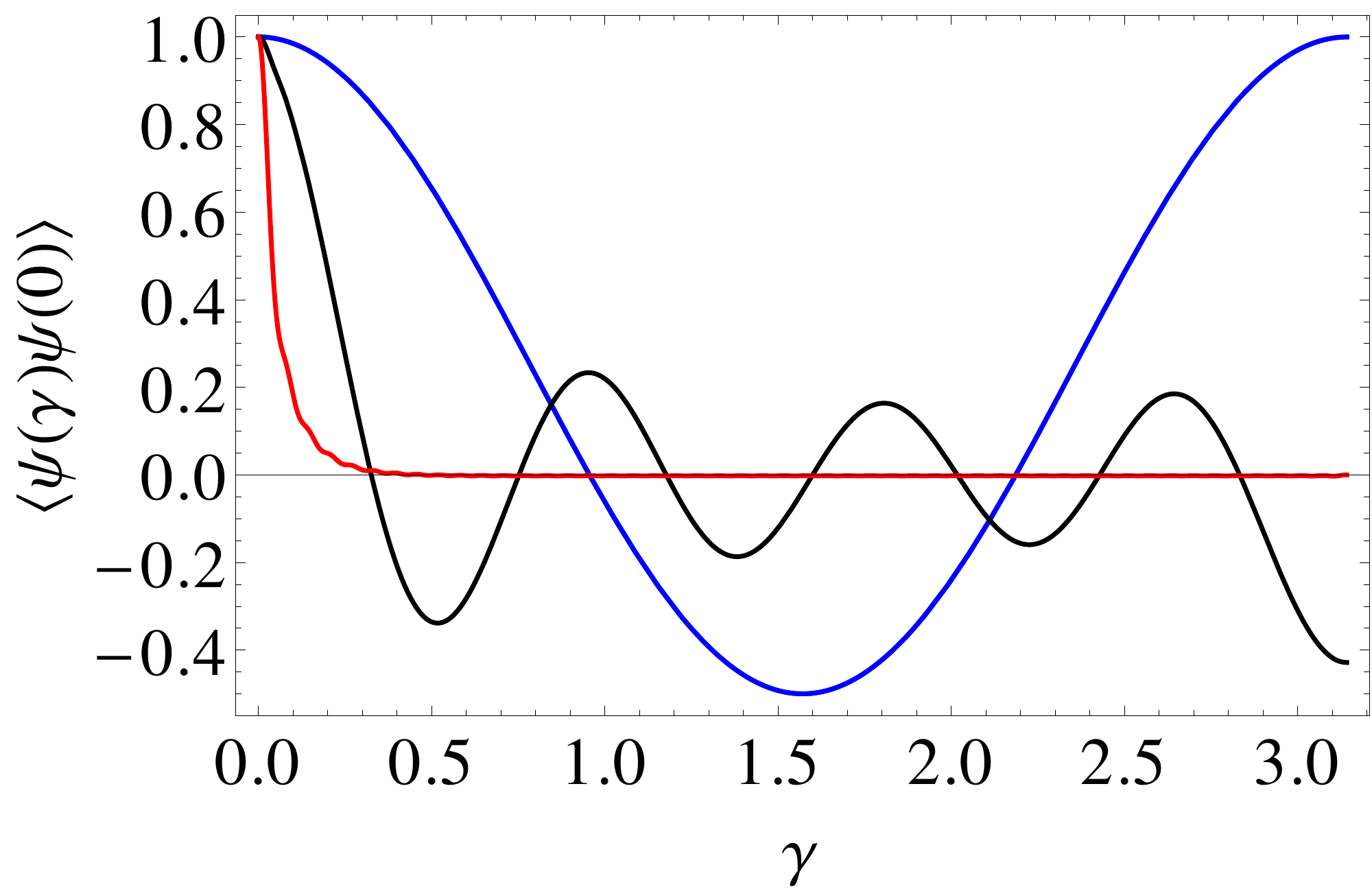}
\caption{\label{fig4} Same as Fig.~\ref{fig3}(a) for $c_1=0$ ($\hat \sigma=25$, $\hat m_+=\hat m_-=10$,  $\hat J=0.05$). In the L phase close to the macrophase transition (blue, $\psi_-$ and $\psi_+$ are superimposed) ($\hat \kappa_1=2$, $c_0=0.9$, $\hat m_0=1.02$) and close to the O one (red for $\psi_+$ and black for $\psi_-$) ($\hat \kappa_1=8.58$, $c_0=0.8$, $m_0=0$).
}
\end{figure}
More interesting are the cases where $\kappa_1\neq0$ since the composition dependent bending modulus plays a role only for spherical membranes, at the Gaussian level. In fig.~\ref{fig4}(a) is shown the correlation functions in the L and O phases, as in  fig.~\ref{fig3}(a), for $c_1=0$ and $\hat \kappa_1=2$ and 8.58. Although the liquid phase is quite similar to the previous case (the two maxima have different values because we are far from the macrophase separation), one observes, \textit{even in the totally decoupled case} $M_0(l)=0$ (\textit{i.e.} $m_0=0$ and $C_1=0$), the signature of an O phase in $\langle \psi_+(\gamma)\psi_+(0)\rangle$, whereas for the same parameter values, $\langle \psi_-(\gamma)\psi_-(0)\rangle$ has a liquid behavior. The system prefers to have highly correlated raft-like domains of large bending rigidity to minimize its free energy. This affect cannot be observed for planar membranes.

\subsection{Phase diagrams}
\begin{figure}[t]
\includegraphics[width=.45\textwidth]{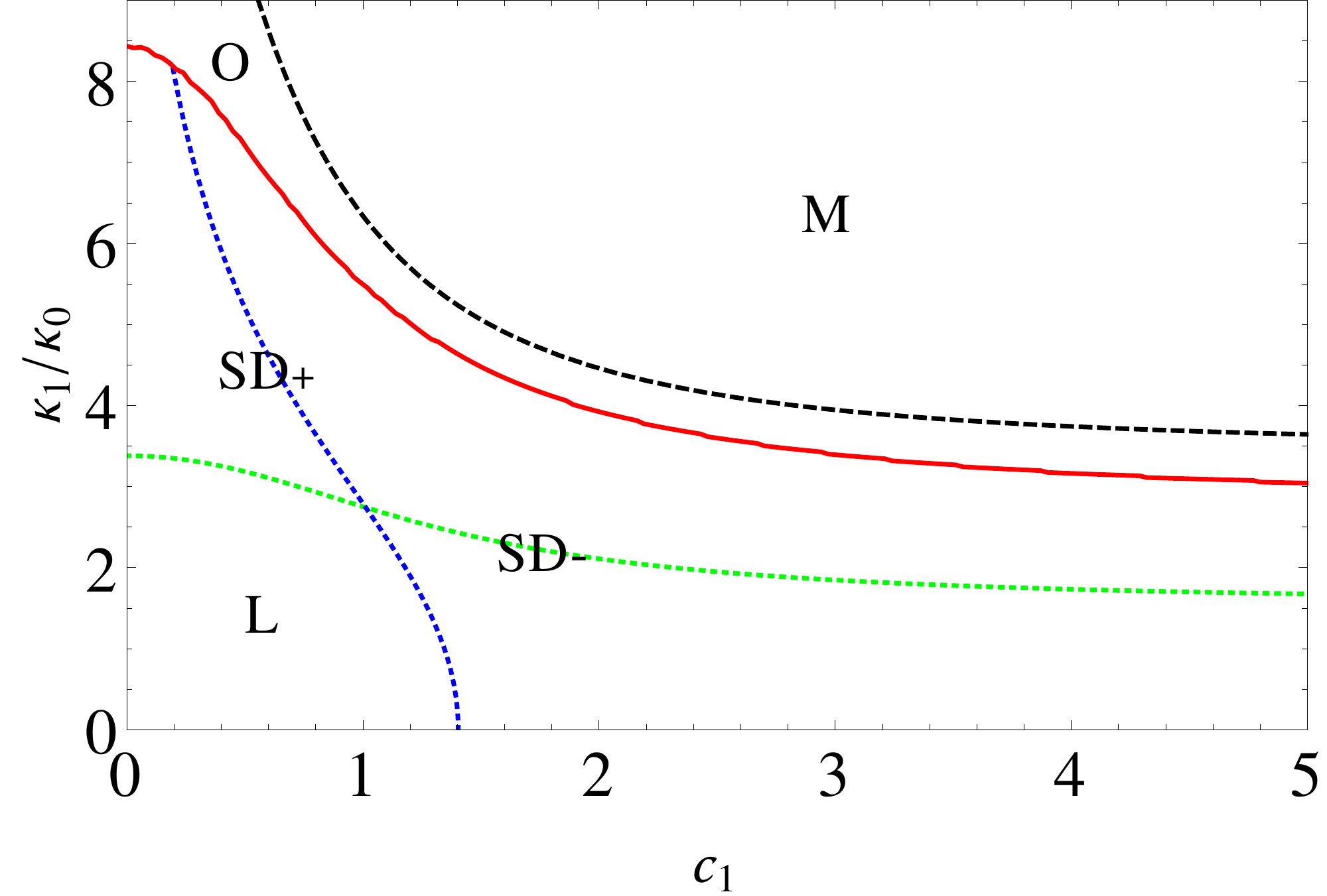}
\caption{\label{fig5} Phase diagram for uncoupled monolayers ($m_0=0$) in spherical membranes for $\hat \sigma =25, \hat m_+=\hat m_-=10, c_0=1.6$, and $\hat J=0.05$. The L phase exists only for low coupling parameters $\hat \kappa_1$ and $c_1$, the SD phases (blue and red dotted lines) set up for intermediate values and the phase transition towards an O phase (red line) occurs before the macrophase separation (dotted black line).}
\end{figure}
In this section, we make the choice to present three different phase diagrams for $\hat m_0=0,5$, and 11.25, which correspond to reasonable values, but all the possible phase diagrams cannot have been explored extensively. Moreover, we are not able to discuss the case of low temperature modulated phases like in~\cite{taniguchi1994}, since we do not have a fourth order term in our Hamiltonian.

A phase diagram is shown for $m_0=0$ in fig.~\ref{fig5}. Note that due to terms in $\hat \kappa_1$ and $c_1$ in \eq{M0}, both leaflets are coupled, which makes the appearance of SD and O phases possible. Figure~\ref{fig5} clearly shows a large region where structured phases emerge, as soon as $C_1\gtrsim C_0$ and $ \kappa_1\gtrsim3\kappa_0$ which are reasonable values. For instance, the theory of elasticity of continuous media shows that the bending modulus of a thin plate varies with its thickness $e$ as $\kappa\propto e^3$~\cite{landau}. Hence an increase of about 40\% of the thickness of each monolayer in the raft domains would be enough to enter the SD regime. More precisely, for $C_1=0$ and $m_0=0$, the SD+ regime occurs for  $\kappa_1\geq\kappa_1^*$, where the critical bending modulus is
\be
\kappa_1^*=\frac{4\sqrt{J\kappa_0}}{(2-c_0)\sqrt{6-c_0}}\frac{6+\hat\sigma-c_0(2-c_0)}{[\hat \sigma(2+c_0)-6(2-c_0)+c_0^3]^{3/2}}
\label{kappa1star}
\ee
Hence it is proportional to the line tension $\propto\sqrt{J}$ as $C_1^*$ in \eq{C1star} and is defined for $\sigma=0$ as soon as the curvature $c_0$ is larger than $1.47$ [the real root of the denominator of \eq{kappa1star}]. This is related to the renormalization of the surface tension by curvature terms, as stressed in Section~II. Besides, \eq{kappa1star} shows that, in the planar case ($c_0\to2$), $\kappa_1^*\to\infty$, \textit{i.e.} the SD+ phase does not appear at $C_1=0$.

\begin{figure}[t]
\includegraphics[width=.45\textwidth]{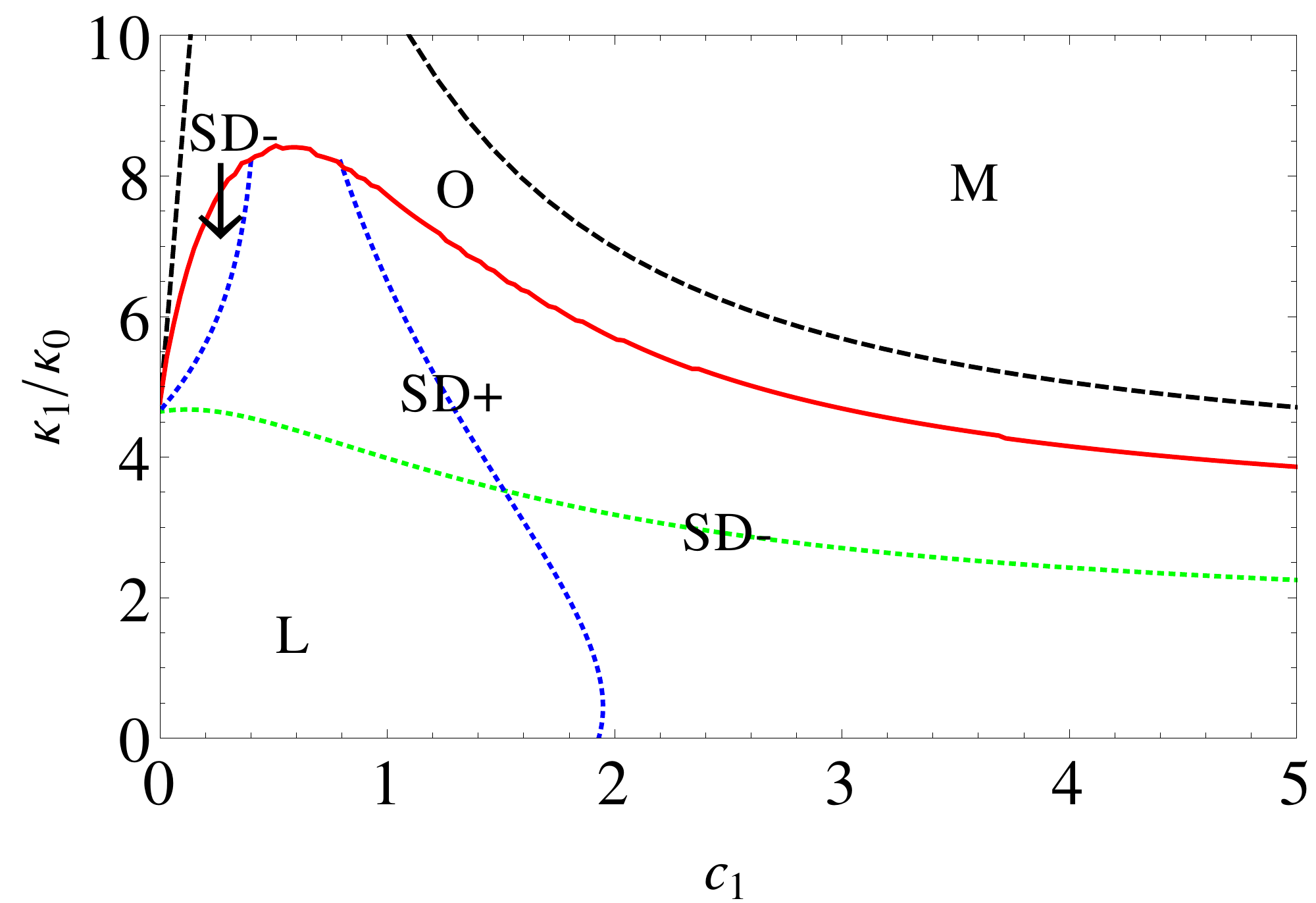}\hfill
\includegraphics[width=.45\textwidth]{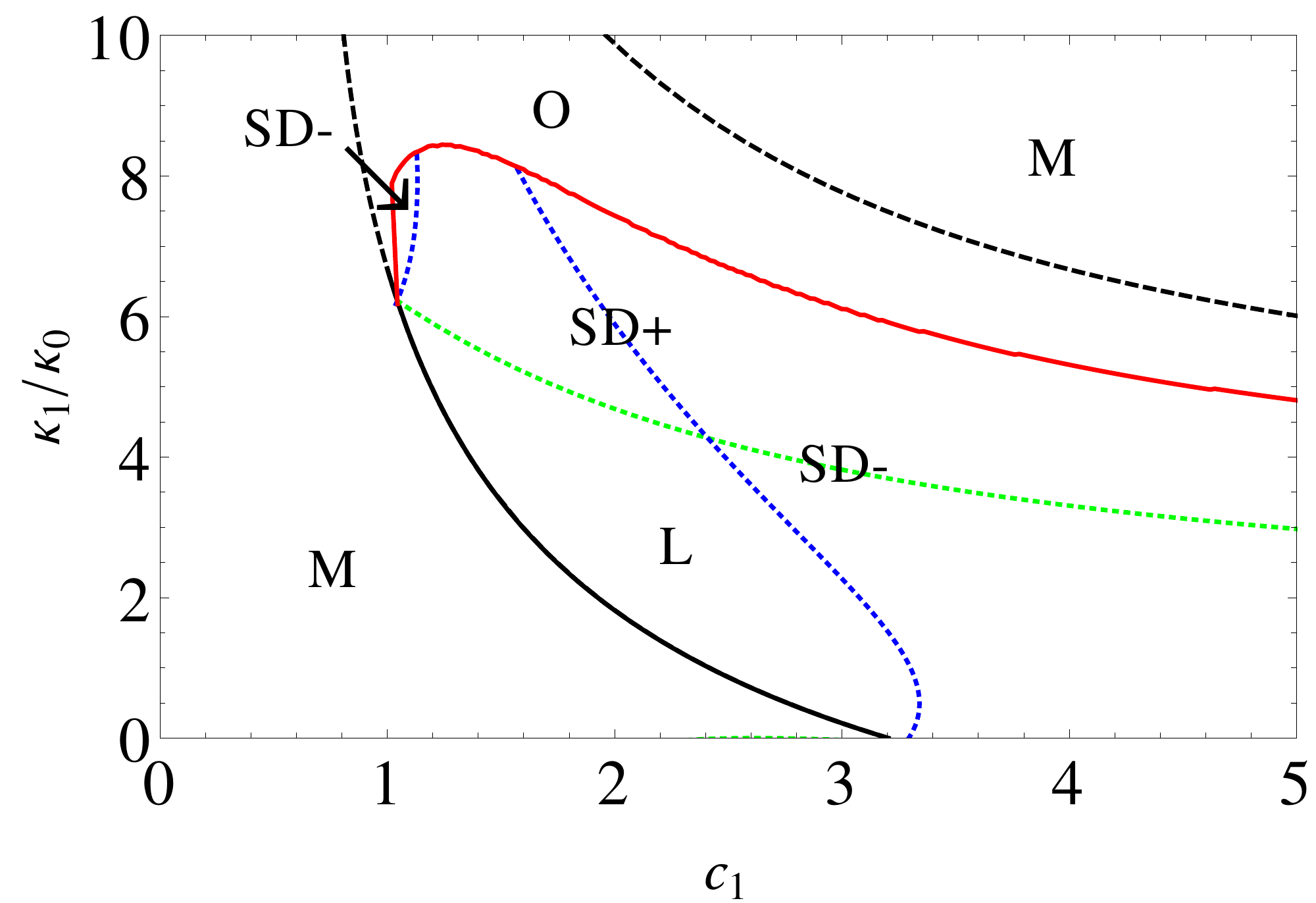}
\caption{\label{fig6} Phase diagram for coupled monolayers ($m_0=5$ and 11.25) in spherical membranes (same parameter values as in fig.~\ref{fig5}). The increase in leaflet coupling leads to phase separation (M) and reduces the L phase region. Note that a tricritical point appears at $(0,\kappa_1^*(m_0))$ for $m_0=5$, which is shifted to large $c_1$ when $m_0$ increases.}
\end{figure}

When $m_0$ is non-zero, the coupling between the two monolayers is reinforced, and the L region in the phase diagram becomes smaller, as shown in fig.~\ref{fig6} for $m_0=5$ and 11.25.
Furthermore the coupling increases slightly the critical bending modulus for $C_1=0$ according to
\be
\kappa_1^*(\hat m_0)=\kappa_1^*\sqrt{1+\left(\frac{\hat m_0}{\hat m_-+6\hat J}\right)^2}
\ee
For $m_0=11.25$, one observes a 2-phases region for small $c_1$ and $\hat \kappa_1$. Surprisingly, the L phase is favored by increasing $c_1$ or $\hat \kappa_1$, which is essentially due to the decrease of $M_0(l)$ due to the term in $-\kappa_1c_1(2-c_0)$ in \eq{M0}.

\section{Discussion}

Due to the large number of parameters, we have plotted typical lipid-lipid correlation functions  for spherical membranes and the corresponding phase diagrams in figs.~\ref{fig3}-\ref{fig6}, which show some general trends.\\ 
\textit{(i)}~Due to the curvature-composition coupling, controlled by $C_1$, and the bending-composition one, controlled by $\kappa_1$, Structured Disordered phase --also called modulated-- phases, showing oscillating correlation functions of $\psi_1-\psi_2$  (SD-) or $\psi_1+\psi_2$ (SD+), emerge from the homogeneous liquid (L) phase.\\
\textit{(ii)}~For the first time to our knowledge, we propose the occurrence of a structured disordered phase of thick and flat patches (SD+ phase) for spherical membranes. This phase is composed of transient (metastable) ``rafts'', and appears for large $\kappa_1>\kappa_1^*$ at $C_1=0$ [see \eq{kappa1star}] and for even lower values of $\kappa_1$ when $C_1>0$. At low coupling between the two leaflets, this new bending-induced mechanism for the formation of rafts is the result of the competition between two forces: to minimize its bending energy, the system prefers to form thick domains (rafts) with large bending modulus, $\kappa_1$ [the mass $m_+$ is renormalized by a term in $\hat \kappa_1^2(2-c_0)^2$, see \eqs{thetaplus}{Mplus}], which is balanced by the usual cost of demixing (line tension), in $\hat J(\tilde\nabla \psi_+)^2$, which favors shorter interfacial lines between rafts and the surrounding phase. Balancing these two terms leads to a critical wave-number $l^*\simeq\kappa_1(2-c_0)/\sqrt{2J}$ in agreement with \eq{kappa1star}.\\
\textit{(iii)}~At large coupling between the leaflets (\textit{i.e.} large $m_0$), which is related to differences between the average concentration $\bar\phi_1$ and $\bar\phi_2$, a structuration made of both rafts and curved patches occurs driven by the high correlation between curved and thick patches. For higher coupling, a true (meso-)phase transition occurs towards an Ordered phase, where the structure factor diverges at a non zero wave number $l^*$. However, since $l^*$ cannot be an integer, this phase transition is actually a crossover. This Ordered phase cannot be studied at the Gaussian level, since high order terms in $\psi^4$ should be properly included to stabilize the membrane.\\ 
\textit{(iv)}~We recover the planar membrane case by taking the limit $c_0\to 2$. The bending-composition coupling disappears at the Gaussian level, and the \textit{structuration is made of curved patches only} (SD- phase), for 
\be
C_1>C_1^*\sqrt{1+(\ell_+/\ell_0)^4}
\ee 
A divergence of the structure factor at $q^*\neq0$ (phase transition towards an Ordered phase) is possible only for large coupling between the leaflets (\textit{i.e.} large $\ell_0^{-2}\propto m_0$). This structured phase can be a stripe phase or an hexagonal one depending on the temperature and the concentrations~\cite{leibler2,sunilkumar,taniguchi1994}.\\
\textit{(v)}~Contrary to previous works, where the rafts were assumed to be curved patches according to the curvature-induced mechanism~\cite{schick,shlomovitz,hirose}, we identify the apparition of rafts, which are known to be flat and thick (and thus stiff)~\cite{simons2,schmid}, with the SD+ phase (an also the O one), which, in the planar case, is only induced by the cross-correlation between curved and stiff patches for large $C_1$ and $m_0$.

The values of the parameters $\kappa_0$ and $\kappa_1$ that we found in the literature are consistent with the critical values observed in the phase diagrams of figs.~\ref{fig5}-\ref{fig6}. Indeed, using micropipette pressurization of giant bilayer vesicles, the bending moduli $\kappa_0$ and $\kappa_1$ and the surface tension can be inferred. By changing the number of unsaturations (from 1 to 6) or the lengths (from 13 to 22 carbons) of diacyl phosphatidylcholine (PC) lipids, Rawicz \textit{et al.}~\cite{rawicz} found extreme values  for $\kappa_+=1.2\times 10^{-19}$~J down to $\kappa_-=0.4\times10^{-19}$~J, which yields $\kappa_1=0.8\times10^{-19}$~J (around $20~k_{\rm B}T$) and thus values of $\kappa_1/\kappa_0$ up to 2 for these lipids.  
Vind-Kezunovic \textit{et al.}~\cite{vind} measured, in unperturbed human HaCaT keratinocytes, composed of a mixture of lipids and cholesterol, values for the bending modulus $\kappa_0=2.7\times 10^{-19}$~J and $\kappa_1\simeq3\,\kappa_0$. 
In intestinal cells, rafts are lipid ordered domains enriched in sphingolipids and cholesterol~\cite{simons1}, since the sphingolipids have a higher affinity with cholesterol than the phospholipids which essentially remain in the disordered phase. One important point in our study is the asymmetry in the leaflet compositions. Usually this difference in lipid average surface fractions are maintained, in living cells, by the Golgi complex~\cite{simons1} and by enzymes named flipases~\cite{lodish}. 
Typical values for $C_1$ can be estimated from giant unilamellar vesicles made of ternary lipid mixtures (sphingomyelin, DOPC and cholesterol), which exhibit curved domains of micrometric radius of curvature~\cite{baumgart}, which leads to values of $c_1=C_1R$ from 0 to 10.\\

Most of the previous works on mixed membranes have been done for planar membranes, and therefore, we focus here essentially on the comparison of our model with these works~\cite{leibler1,leibler2,mackintosh,schick,sunilkumar,shlomovitz,hirose}. First, a constant spontaneous curvature $C_0\neq0$ is not compatible with the planar geometry at large length scales. The up/down symmetry can only be broken locally. Hence these works focussed on the $C_0=0$ case, which we have seen to be very restrictive.
Second, as already said above, only the curvature-induced mechanism subsists for planar membrane. In almost all these studies, the coupling is introduced directly with a term in $-\kappa_0C_1\int \psi_-\nabla^2 h $, contrary to a more natural hamiltonian $\frac12 \kappa_0\int (\nabla^2 h -C_1\psi_-)^2$. The direct and unexpected consequence is that, in the last case, the prefactor of $\psi_-^2$ (the mass of the field theory) is renormalized by a term in $\kappa_0C_1^2$ which forbids the formation of mesophases a low coupling between $\psi_-$ and $\psi_+$ [fig.~\ref{fig2}(b)]. Hence our phase diagram is different from the ones obtained in refs.~\cite{leibler1,mackintosh,schick,sunilkumar,hirose}. Furthermore, in some of these works~\cite{leibler1,mackintosh,sunilkumar} a confusion is made between the structured disorder phase and the ordered one. Indeed, phase diagrams are drawn by comparing the free energies of the liquid phase and the structured disordered by selecting only $q=0$ and $q^*$ in the structure factors respectively (saddle point approximation), which is equivalent to not consider the damping of the correlation functions show in fig.~\ref{fig2}(c).
Hirose \textit{et al.} and Schick pointed out this confusion recently~\cite{schick,hirose}. We find the same order of magnitude of $\xi$ for the characteristic correlation lengths and wave vectors. Note however that we cannot compute analytically the poles of the structure factors, contrary to~\cite{schick,hirose} where the calculation of the structure factors was limited to small wave-vectors $q<\xi^{-1}$. This is of course not verified for large $C_1$.

In refs.~\cite{shlomovitz,hirose}, the coupling is introduced directly between the lipid compositions of the two monolayers, which corresponds in our notations to a term in $-k\psi_1\psi_2$, whereas the coupling with the membrane height fluctuations is done separately on the two monolayers (only on the upper one in~\cite{shlomovitz}). Then the two  height fluctuation fields $h_1$ and $h_2$ are integrated out separately. In our model, however, we keep $h$ defined in the middle plane of the bilayer and we suppose the membrane to be infinitely thin, as usually done in elasticity~\cite{landau}. It allows us to define properly rafts as thick patches which are \textit{not curved}, as seen in simulations~\cite{stevens,schmid}, contrary to the models of refs.~\cite{schick,shlomovitz} where the rafts are necessarily curved.

Meinhardt \textit{et al.} have numerically addressed the issue of the correlation between liquid ordered domains in opposed monolayers~\cite{schmid}. They observed that when increasing the concentration in lipids of type A, presumably in both monolayers, the cross-correlation between liquid ordered domains becomes significant, which can be translated, in our model, by an emergence of thick patches. This is consistent with our model since increasing the concentration in lipids of type A in both leaflets leads to the decrease of the masses $m_\pm$ (provided that it remains $<1/2$) though $\kappa_0$ increases. It thus corresponds to a decrease of $\ell_\pm^{-1}$ and $\xi^{-1}$, which, from \eq{LSD+} and for a given $m_0$, favors the SD+ phase.

Very recent simulations showing modulated phase patterns on the surface of giant unilamellar vesicles, underline the role played by a difference in bending~\cite{amazon,amazon2014} or/and Gaussian moduli~\cite{hu2011,kumar1997} of the different lipid domains. They assumed to be in the Ordered phase, and showed that the observed patterns are similar to those observed in experiments. As in our model, the morphology of these domains results from the competition between the line tension and the bending energy. However, they did not consider any local spontaneous curvature $C_1$. Note that we do not consider the Gaussian bending rigidity in our model, since we limit our discussion to deformations that preserve the topology of the membrane (and the Gaussian bending energy is constant). However, when finite domains appear this Gaussian rigidity might play a role~\cite{amazon,amazon2014}.

This work can be further developed in several directions. First, by going beyond the Gaussian theory by performing a cumulant expansion~\cite{dean}, we will introduce a coupling term in $\kappa_1$ for planar membranes. This is a work in progress. Note that, field-theoretic methods have been used to treat an assembly of inclusions that both modify the bending rigidity modulus and the local spontaneous curvature for given inclusion distributions~\cite{netz}. Second, by introducing a different coupling between the two monolayers,  the case of asymmetric monolayers with different lipids could be explored. Furthermore, the model can be refined by introducing diagonal inter-plane interactions following~\cite {sunilkumar}. Preliminary studies suggest that the structure disordered and ordered phase are favored. 
Finally, an interesting development of this study would be to explore the role of stiff patches in the budding of fluid membranes~\cite{lipowsky,julicher,harden,ladji2004,kumar2000}. To do so, the potentials $V(\psi)$ should be extended up to $\psi^4$ allowing us to compute the line tension between stiff or curved ordered domains and disordered ones. The line tension of the curved domain edge is known to play a crucial role in budding~\cite{lipowsky}.

\appendix

\section{Chemical potential}
\label{appB}

In writing \eq{H0}, we do not impose that the minimum of the free energy of the system is for $\bar{\phi}_1$ and $\bar{\phi}_2$, and we assume that $\bar{\phi}_1$ and $\bar{\phi}_2$ are imposed by the cell itself or, in the vesicle context, that we are in an out-of-equilibrium state following the vesicle formation. But an alternative possibility is to begin with a more general form of the Hamiltonian~\cite{leibler2,hirose,taniguchi1994}
\begin{eqnarray}
{\cal H}[\phi_1,\phi_2,u] = \int_\mathcal{A} \mathrm{d}\mathcal{A}  \left[\frac{J}2 g^{ij}\nabla_i\phi_1\nabla_j \phi_1 +\mu_1\phi_1+ \frac{m_1}2 \phi_1^2 \right. 
\nonumber \\  +\left.\frac{J}2 g^{ij}\nabla_i\phi_2\nabla_j \phi_2 +\mu_2\phi_2+ \frac{m_2}2 \phi_2^2 + \frac{k}2 (\phi_1-\phi_2)^2\right] \nonumber \\ 
+\sigma \mathcal{A} + \frac12\int_\mathcal{A} \mathrm{d}\mathcal{A}  \,\kappa(\phi_1+\phi_2)\left[\mathrm{div}(\bold n) - C(\phi_1-\phi_2)\right]^2 
\end{eqnarray}
and minimizing it a the mean-field level to compute the average leaflet compositions $\bar{\phi}_i$.
 
It is still possible to use $\phi_+=(\phi_1+\phi_2)/2$ and $\phi_-=(\phi_1-\phi_2)/2$ instead of $\phi_1$ and $\phi_2$ so that the Ginzburg-Landau Hamiltonian becomes
\begin{eqnarray}
{\cal H}_{\rm GL}[\phi_+,\phi_-] = \kappa_0 \int_\mathcal{S} \mathrm{d}\Omega \left[\hat J(\tilde{\nabla}\phi_+)^2+\hat J(\tilde{\nabla}\phi_-)^2\right. \nonumber\nonumber\\\left.+ \tilde\mu_+\phi_+ + \tilde\mu_-\phi_- + \frac{\hat m_+}2\phi_+^2+ \frac{ \tilde m_-}2 \phi_-^2 + \tilde m_0 \phi_+\phi_-\right]
\end{eqnarray}
By writing $\phi_{\pm}(\Omega)=\bar{\phi}_{\pm}+\psi_{\pm}(\Omega)$,  we obtain \eq{GL}, but with $\tilde \mu_+=\hat\mu_++\hat \kappa_1 (2-c_0)^2$, $\tilde \mu_- = \hat\mu_-- c_1(2-c_0)$, (same convention as in Section~\ref{model} for dimensionless parameters) where $\hat\mu_-= \hat\mu_1-\hat\mu_2$ and $\hat\mu_+ =\hat\mu_1+\hat\mu_2$.  The two mean-field equations that fixe $\bar{\phi}_{\pm}$ are then: 
\begin{eqnarray}
\tilde{\mu}_++\tilde{m}_+\bar{\phi}_++\tilde{m}_0\bar{\phi}_-=0 \\
\tilde{\mu}_-+\tilde{m}_-\bar{\phi}_-+\tilde{m}_0\bar{\phi}_+=0
\end{eqnarray}
The important result is that our effective Hamiltonian is unchanged. However, contrary to the case of given values for the average compositions $\bar{\phi_1}$ and $\bar{\phi_2}$, they are now necessarily the positions of the minimum of $\cal{H}_{\rm GL}$~\cite{sunilkumar}.

\section{Curvature for planar limit}
\label{appA}

In this appendix, it is shown why, in the limit of planar membranes, $R \to\infty$, we have to impose $c_0=C_0R=2$.
For planar single component membranes, the normalized excess area is at leading order:
\begin{eqnarray}
\frac{\cal A}{\mathcal{A}_p} &=&1+\frac{1}{2L^2}\int_L {\rm d}{\bold x}^2  \langle (\nabla h)^2\rangle \nonumber \\ 
&=&1+\frac{k_{\rm B} T}2\int_0^{\Lambda} \frac{{\rm d}{\bold q}^2}{(2\pi)^2}\ \frac{q^2}{\kappa_0 q^4+\sigma q^2} \nonumber \\
&=& 1+\frac{k_{\rm B} T}{8\pi\kappa_0}\ln \left(1+\frac{\Lambda^2\kappa_0}{\sigma}\right) \label{Aflat}
\end{eqnarray}
In the spherical case, using \eq{area}, one writes a similar expression~:
\begin{equation}
\frac{\cal A}{\mathcal{A}_p} = 1+\frac{k_{\rm B} T}{8 \pi \kappa_0} \ \sum_{l=2}^{\Lambda R} \ \frac{2l+1}{l(l+1)+\frac{\sigma R^2}{\kappa_0}-c_0(2-\frac{c_0}{2})} \nonumber
\end{equation}

In the limit of large vesicles, $R \to\infty$, we replace the discrete modes of spherical harmonics $l$ by a continuum over $q$ (Sec 4.A), $\displaystyle{\lim\limits_{R \to \infty} l(l+1)/R^2=q^2}$, which allows us to rewrite ${\cal A}/\mathcal{A}_p$. However, to properly take the mode $q=0$ into account, the mode $l=1$ (related to the translation of the vesicle) must be included in the sum, which enforces the choice $c_0\to2$ (so that $l(l+1)-c_0(2-c_0/2)=0$ for $l=1$). Then we simply get:
\be
\frac{\cal A}{\mathcal{A}_p}  = 1+\frac{k_{\rm B} T}{8 \pi \kappa_0} \ \int_{q=0}^{\Lambda } \ \frac{2q{\rm d}q}{q^2+\frac{\sigma }{\kappa_0}} 
\ee
which yields the same result as \eq{Aflat}.

\end{document}